\documentclass[12pt,preprint]{aastex}

\usepackage{graphicx}
\usepackage{txfonts}
\begin{document}

\title{Bayesian Magnetohydrodynamic Seismology of Coronal Loops}

\shorttitle{Bayesian Magnetohydrodynamic Seismology}

\author{I. Arregui\altaffilmark{1} \and A. Asensio Ramos\altaffilmark{2,3}}
\altaffiltext{1}{Departament de F\'isica, Universitat de les Illes Balears,
E-07122, Palma de Mallorca, Spain}
\altaffiltext{2}{Instituto de Astrof\'{\i}sica de Canarias, E-38205, La Laguna, Tenerife, Spain}
\altaffiltext{3}{Departamento de Astrof\'{\i}sica, Universidad de La Laguna, E-38205 La Laguna, Tenerife, Spain}

 \email{inigo.arregui@uib.es, aasensio@iac.es}

\begin{abstract}
We perform a Bayesian parameter inference in the context of resonantly damped transverse coronal loop oscillations. 
The forward problem is solved in terms of parametric results for kink waves in one-dimensional flux tubes in the thin tube 
and thin boundary approximations. For the inverse problem, we adopt a Bayesian approach to infer the most probable 
values of the relevant parameters, for given observed periods and damping times, and to extract their confidence levels. 
The posterior probability distribution functions are obtained by means of Markov Chain Monte Carlo simulations, incorporating 
observed uncertainties in a consistent manner. We find well localized solutions in the posterior probability distribution 
functions for two of the three parameters of interest, namely the Alfv\'en travel time and the transverse inhomogeneity length-scale. 
The obtained estimates for the Alfv\'en travel time are consistent with previous inversion results, but the method enables us to 
additionally constrain the transverse inhomogeneity length-scale and to estimate real error bars for each parameter. 
When observational estimates for the density contrast are used, the method enables us to fully constrain the three parameters of interest.
These results can serve to improve our current estimates of unknown physical parameters in coronal loops and to test 
the assumed theoretical model. 
\end{abstract}



\section{Introduction}\label{intro}

Magnetohydrodynamic (MHD) seismology stands as one of the few indirect methods for the determination of difficult to 
measure physical parameters in solar coronal structures. It relies on the combined use of observed oscillatory properties 
of MHD waves in solar atmospheric magnetic and plasma structures together with theoretically obtained wave properties.  
It was first suggested by  \citet{uchida70} and \cite{REB84}, in the coronal context, and by \citet{TH95}, in the context of solar 
prominences. The last years increase in the number and quality of observations of wave activity in the solar atmosphere and the 
refinement of  theoretical models have allowed the practical implementation of this technique.  In the context of coronal loops, coronal 
seismology has allowed the estimation and/or restriction of  relevant parameters such as the magnetic field strength 
\citep{Nakariakov01}, the Alfv\'en speed \citep{temury03,Arregui07,GABW08}, the transversal density structuring \citep{verwichte06},  
or the coronal density scale height \citep{AAG05,Verth08}.  Of particular relevance has been the use of the concept of period 
ratios as a seismological tool, first pointed out by  \cite{AAG05,GAA06}, and reviewed by \cite{andries09}.

Most of the efforts in this area have been concentrated on the phenomenon of quickly damped transverse oscillations in 
coronal loops, first reported by \citet{Aschwanden99} and \cite{Nakariakov99}. These oscillations are interpreted in terms of linear 
MHD kink waves of a magnetic flux tube, a wave mode with mixed fast and Alfv\'en character, its Alfv\'enic nature being dominant in 
and around the resonant position \citep{Goossens09}, where the global eigenmode frequency matches the local Alfv\'en frequency. 
Starting with the simplest model that considers the fundamental transverse oscillation of a magnetic flux tube \citep{ER83} several 
model improvements have included other effects, such as the curvature of coronal loops, density stratification, or the departure from circular 
cross section of the tubes \citep[see][for a recent review]{Ruderman09}. All these new ingredients have been seen to produce second 
order effects on the main wave properties.

The advancement in seismology of kink modes in coronal loops is reviewed by \cite{Goossens08}. 
\cite{Nakariakov01} used the observed periods and theoretical estimates of the periods, based on the long
wavelength  approximation  for a uniform coronal loop model, to derive estimates for the magnetic field strength, by making assumptions on 
the density. \cite{GAA02} used the observed damping rates and theoretical values of the damping rates, based on the thin boundary 
approximation, to derive estimates for the radial inhomogeneity length-scale, once the density contrast was assumed. \cite{Aschwanden03b} used the 
observed damping rates and the damping rates computed by \cite{tom04b}, outside 
the thin boundary approximation, to compare the theoretically predicted density contrasts to estimates obtained indirectly taking into account effects such as the subtraction 
of the background flux, the line-of-sight integration of the emission measure, the spatial smearing due to the transverse motion, and the point-spread function of the instrument. In their study, the internal density could only be determined as a function of the external density.

The main limitation of these studies is that the determination of the magnetic field strength \citep{Nakariakov01} or transverse inhomogeneity length-scale \citep{GAA02}
is only possible if a given value for the density contrast is adopted. Otherwise, an infinite number of solutions to the inverse problem arise. 
This was shown in the numerical and analytic seismological inversions by \cite{Arregui07} and \cite{GABW08},  that combine the observational information on both periods and damping times for resonantly damped kink modes in a consistent manner.  These inversions enable to infer  
information about  both the internal Alfv\'en speed and the transverse density structuring.  Their approach is to make no assumption on the particular value 
of any of  the physical parameters of interest. Their ranges of variation, compatible with observations, are instead obtained \citep{GABW08}. Following this approach,  
a complete solution to the inverse problem is obtained, from which general properties and liming cases can be studied. 
The solution gives rise to an infinite number of equally valid equilibrium models 
that explain observations, characterized by three parameters: the density contrast, the transverse inhomogeneity length-scale, and the Alfv\'en speed. In addition, in the inversion 
techniques presented by \citet{Arregui07} and \citet{GABW08}, it is not straightforward to devise a consistent method to compute uncertainties on the inferred parameters 
from the measurement errors on observed periods and damping rates.

Bayesian inversion techniques for parameter inference can help us to overcome these limitations and are applied in this paper to
the determination of physical parameters in oscillating coronal loops. These techniques have proven to be useful in a number
of areas of physical sciences \citep{Gregory05}, including solar physics. In the context of 
inversion of Stokes profiles, \cite{andres07} apply Bayesian techniques to analyze the performance of a given radiative transfer model 
to fit a given observed Stokes vector aiming at obtaining information about the thermodynamic and magnetic properties of solar and stellar 
atmospheres \citep[see also][]{andres11}. \cite{Marsh08} use a Bayesian probability-based approach to the problem of detecting and parameterizing oscillations in the 
upper solar atmosphere, in order to determine the number of oscillations present, and their properties.  In the context of coronal heating,  
\citet{adamakis10} employ Bayesian model comparison techniques  to 
determine the preferred heating location along coronal loops. Finally, \cite{Ireland10} have recently presented an automated detection 
technique for oscillating regions in the solar atmosphere based on  Bayesian spectral analysis of time series and image filtering. 

In this paper, a Bayesian MHD seismology inversion technique is presented and the results from its application to coronal loops are described.
The layout of the paper is as follows. Section~\ref{forinv} describes the physical model 
and the forward and inverse problems.  In Section~\ref{tech}, the details of the developed Bayesian technique are given. 
The analysis and results of its application to synthetic data and to real coronal loop oscillations are discussed in Section~\ref{results}.
Finally, in Section~\ref{conclusions}, our conclusions are presented.

\section{The forward and inverse problems}\label{forinv}

The application of the Bayesian formalism to parameter inference is rather general and can be applied to any model  that explains a 
given set of observations.  In our case the theoretical model is the resonantly damped MHD kink mode interpretation of quickly damped 
transverse oscillations of coronal loops. The classic theoretical model assumes that coronal loops  can be represented as straight 
cylindrically symmetric magnetic flux tubes with a uniform magnetic field pointing along the axis of the tube. In the zero plasma-$\beta$ 
approximation coronal loops  are density enhancements with a constant internal density, $\rho_{\rm i}$,  a constant external 
density, $\rho_{\rm e}<\rho_{\rm i}$, and a non-uniform transitional layer of thickness $l$ that connects both regions.  Analytical theory for 
linear MHD kink oscillations based on the thin tube and thin boundary approximations gives us two equations \citep[see][]{GABW08} for 
the period and damping time of resonantly damped oscillations,

\begin{equation}
P  =  \tau_{\rm Ai} \; \sqrt{2} \; \; \left(\frac{\zeta + 1}{\zeta}\right)^{1/2} \label{T} \mbox{\hspace{.5cm}} \mbox{and} \mbox{\hspace{.5cm}}  \frac{\displaystyle \tau_{\mathrm d}}{\displaystyle P} = \frac
{\displaystyle 2} {\displaystyle \pi} \frac{\displaystyle \zeta +
1}{\displaystyle \zeta - 1} \frac{\displaystyle 1}{\displaystyle
l/R}.\label{forward}
\end{equation}
The factor $2/\pi$ in the damping rate expression has its origin on  the assumption of a sinusoidal variation of density across the 
non-uniform layer. These equations express the period, $P$, and the damping time, $\tau_{\rm d}$, which are observable quantities 
in terms of the internal Alfv\'en travel time, $\tau_{\rm Ai}$, the density contrast, $\zeta=\rho_{\rm i}/\rho_{\rm e}$, and the transverse 
inhomogeneity length scale, $l/R$, in units of the radius of the loop. These three quantities ($\tau_{\rm Ai}, \zeta, $ and $l/R$) are the 
seismic quantities in the sense that they are the quantities that we aim to determine with the use of observed values for the period and 
the damping time. Since we only have two equations that relate the three unknown quantities to the two observed quantities there is an 
infinite number of solutions to the inverse problem, as first pointed out by \cite{Arregui07}. These solutions have to follow a precise one-dimensional solution curve in the three-dimensional parameter space of unknowns. 
This curve constitutes a characteristic attribute of resonant absorption subject to observational testing. It was first obtained numerically 
by \citet{Arregui07} and subsequently analytically by \citet{GABW08} under the thin tube and thin boundary approximations. 

Because of the very good agreement between the analytic and numerical inversion solution curves demonstrated in \cite{GABW08}, in 
the following we adopt  the analytic approximation for the forward problem. The clear advantage is that the 
forward  problem becomes algebraic and this simplifies the statistical inversion scheme considerably. For a particular observed event,  
with fixed period and damping time, the seismic variables are constrained to the following intervals

\begin{eqnarray}
\zeta &\; \in \; & I_{\zeta} = [\frac{\displaystyle C +
1}{\displaystyle C - 1}, \;\;\infty[ \nonumber \\
y=\frac{\tau_{\mathrm Ai }}{P} & \; \in \; & I_y = [\frac{\displaystyle 1}{\displaystyle 2} \left
(\frac{\displaystyle C + 1}{\displaystyle C} \right )^{1/2} , \;\;
\frac{\displaystyle 1}{\displaystyle \sqrt{2}}[ \nonumber \\
z =\frac{l}{2R}& \;\in \; & I_z = ]\frac{\displaystyle 1}{\displaystyle C},
\;\;1]
\label{yzez1}
\end{eqnarray} 
with $C=\pi\tau_{\rm d}/P$ known from observations. The seismic variables are not independent, but are related to one another by 
the following functions defined in \cite{GABW08}

\begin{eqnarray}
y&=&F_1(\zeta)=\frac{1}{\sqrt{2}}\left(\frac{\zeta}{\zeta+1}\right)^{1/2}, \hspace{0.2cm}\zeta=G_1(y)=\frac{2y^2}{1-2y^2},\nonumber\\\nonumber\\
z&=&F_2(\zeta)=\frac{1}{C}\frac{\zeta+1}{\zeta-1}, \hspace{0.2cm}\zeta=G_2(z)=\frac{Cz+1}{Cz-1}.\nonumber
\end{eqnarray}
Of the four functions only two  are independent.  This allows to consider one of the three unknowns as a parameter, say $z=l/2R$, let it 
take on all values in $I_{\rm z}$ and then compute the corresponding values of $y=F_1(G_2(z))$ and $\zeta=G_2(z)$. An example of the 
resulting inversion curve is shown in Figure~1 in \cite{GABW08} (see also Figure~\ref{fig:3dchain-priors} in this paper). 

Although all the possible combinations of parameters along the seismic curve equally well explain observations it turns out that the 
obtained curve gives rather constrained values for the Alfv\'en travel time. Moreover, when applied to prominence thread transverse 
oscillations, given their large density contrast values, the asymptotic behavior of the solution curve towards large density contrast 
values allows to derive asymptotic values for the internal Alfv\'en speed and the transverse inhomogeneity length scale \citep{AB10}. 
In spite of these virtues, the problem with the infinite number of solutions remains.  A genuine measurement of any of the three 
unknowns would enable us to collapse the one-dimensional inversion curve into a zero-dimensional solution point. 
The matter is to obtain reliable observational estimates for any of the three unknowns.  Of related importance is  to devise a 
consistent procedure for incorporating to the inversion schemes measurement errors in periods 
and damping times  to compute the uncertainties in the inferred physical parameters.

\section{Probability model and numerical method}\label{tech}

\subsection{Mathematical formulation}

Our model describing resonantly damped transverse loop oscillations contains three free parameters, 
$\zeta$, $l/R$, and $\tau_{\rm Ai}$ that have to be estimated. Let these three parameters be gathered in the vector 
{\boldmath $ \theta$}=[$\zeta$, $l/R$, $\tau_{\rm Ai}$]$^t$ and let the set $d$=$[P_{\rm obs}, \tau_{\rm d, obs}]$$^t$ contain discretized data on period and damping times. When data, as a result 
of measurements determine the values of $d$, the state of knowledge on {\boldmath $\theta$} is represented by 
$p$({\boldmath $\theta$}$| d$), which is given by Bayes' rule \citep{bayes63}

\begin{equation}
p(\mbox{{\boldmath $\theta$}} | d)=\frac{p(d | \mbox{{\boldmath $\theta$}})p(\mbox{{\boldmath $\theta$}})}{p(d)}.\label{bayes}
\end{equation}
In this expression,  $p$({\boldmath $\theta$}$| d$) is the so-called posterior probability distribution. 
The function $p$($d |${\boldmath $\theta$}) is the conditional probability distribution of the data given the parameters. It contains the 
theoretical relations between parameters and data (including the noise properties) and is a measure of how well 
the data are predicted by the model. Before $d$ has been observed, it represents the probability distribution associated with possible 
data realizations for a fixed parameter vector. A posteriori, after observation,  $p$($d |${\boldmath $\theta$}) has a 
very different interpretation. It is the likelihood of obtaining a realization actually observed as a function of the 
parameter vector, and is hence called the likelihood function. Under the assumption that observations are corrupted 
with Gaussian noise and that they are statistically independent, the likelihood can be expressed as

\begin{equation}
p(d|\mbox{\boldmath $\theta$}) = \left(2\pi \sigma_P \sigma_\tau\right)^{-1} \exp \left\{ \frac{\left[P-P^\mathrm{syn}(\mbox{\boldmath $\theta$})\right]^2}{2 \sigma_P^2} +
 \frac{\left[\tau_d-\tau_d^\mathrm{syn}(\mbox{\boldmath $\theta$})\right]^2}{2 \sigma_\tau^2} \right\},
\end{equation}
with $P^\mathrm{syn}(\mbox{\boldmath $\theta$})$ and $\tau_d^\mathrm{syn}(\mbox{\boldmath $\theta$})$  given by 
the forward analytical problem of Equation~(\ref{forward}). Likewise, $\sigma_P^2$ and $\sigma_\tau^2$ are the variances 
associated to the period and damping times, respectively.

The quantity $p$({\boldmath $\theta$}) is the prior probability of the model vector. It should contain any prior information 
we might have on the model parameters,  without taking into account the observed data. Three different versions are used in this paper. The normalization constant on 
the right-hand side of  Equation~(\ref{bayes}) is the marginal likelihood, called ``Bayesian evidence'' in cosmology 
applications. The evidence can be written as

\begin{equation}
p(d)=\int p(d | \mbox{{\boldmath $\theta$}})p(\mbox{{\boldmath $\theta$}}) d\theta,
\end{equation}
which is an integral of the likelihood over the prior distribution that normalizes the likelihood and turns it into a probability.
This quantity is central for model comparison purposes, since it evaluates the model's performance in the light of data. In 
Bayesian inference, to compute the posterior distribution one needs to run the model over the entire parameter space 
of interest and sum that all up.

In the Bayesian formulation the inference of a given parameter set {\boldmath $\theta$} is based on the knowledge of 
$p$({\boldmath $\theta$}$| d$), which contains all the information available on {\boldmath $\theta$} given the data $d$ 
and therefore is the solution to the inverse problem. Once the posterior distribution  $p$({\boldmath $\theta$}$|d$) is 
known, the position of the maximum value gives the most probable combination of parameters  {\boldmath $\theta$} 
that fit the data. Bayes' rule constitutes a simple mathematical formulation of the process of learning from experience. 
What can be inferred about the model parameters ``a posteriori'' is a combination of what is known ``a priori'', independently 
of the data, and of the information contained in the data. It is an appealing formulation that shows how previous knowledge 
can be updated when new information becomes available. An additional advantage of the Bayesian approach is that error
propagation is consistently taken into account by marginalization. When one needs to know how the data $d$
{\bf constrains} one of the parameters, say $\theta_i$, it is enough to compute the following integral

\begin{equation}
p(\theta_i|d) = \int p(\mbox{{\boldmath $\theta$}} | d) d\theta_1 \ldots
d\theta_{i-1} d\theta_{i+1} \ldots d\theta_{N}.
\end{equation}
The result is known as the marginal posterior, which encodes all information for model parameter $\theta_i$ available in 
the priors and the data. Furthermore, it correctly propagates uncertainties in the rest of parameters to the one of interest.

\subsection{Prior information}\label{priors}

The prior choice, $p$({\boldmath $\theta$}), is a fundamental ingredient of Bayesian statistics, in order to obtain optimal results. The results obtained from 
the statistical inference should be independent on prior information, provided the prior has a support that is non-zero in regions 
of the parameter space where the likelihood is large. In this case, repeated application of Bayes' rule will lead to a 
posterior probability distribution that converges to a common objective inference on the hypothesis \citep{trotta08}. Two scientists in the 
same state of knowledge should assign the same prior, hence the posterior should be identical if they observe the same data. 
In our particular case, dealing with the inversion of physical parameters in
oscillating coronal loops, expressions~(\ref{yzez1}) define the intervals over which the unknown physical parameters are allowed to
vary, according to the analytic inversion scheme by \cite{GABW08}.

When all  information we have on unknown parameters is restricted to ranges of variation, a reasonable choice is to
assign the same probability to all values contained within that range. This defines a uniform type prior, and we can write

\begin{equation}\label{uniform}
p(\theta_i)=H(\theta_i, \theta^{\mathrm{min}}_i, \theta^{\mathrm{max}}_i),
\end{equation}
where $H(x, a, b)$ is the top-hat function

\begin{equation}
H(x, a, b)=\left\{\begin{array}{ll}
\frac{1}{b-a}\hspace{0.8cm} \textrm{$a\le x \le b$},\\\\
0  \hspace{1.1cm} \textrm{otherwise}.\\
\end{array}\right.
\end{equation}

Another option is to assign a decreasing  probability distribution for increasing values of the parameter, over a given range. 
This can be accomplished by considering a Jeffreys' type prior, and we can write

\begin{equation}\label{jef}
p(\theta_i) = \left[\theta_i \log{\left(\frac{ \theta^{\mathrm{max}}_i}{\theta^{\mathrm{min}}_i}\right)}\right]^{-1}.
\end{equation}

If some additional information about the unknown parameter is available from observations, a  Gaussian distribution centered on the measured value can be used, so that

\begin{equation}\label{gaussian}
p(\theta_i) = \left(2\pi \sigma_{\theta_{i}}^2\right)^{-1/2} \exp \left[ \frac{-\left(\theta_i-\mu_{\theta_i}\right)^2}{2 \sigma_{\theta_i}^2}  \right].
\end{equation}
In this expression, the mean $\mu_{\theta_i}$ and the standard deviation $\sigma_{\theta_i}$ would be directly  obtained from observations.

\begin{figure}[t]
   \centering
  \plotone{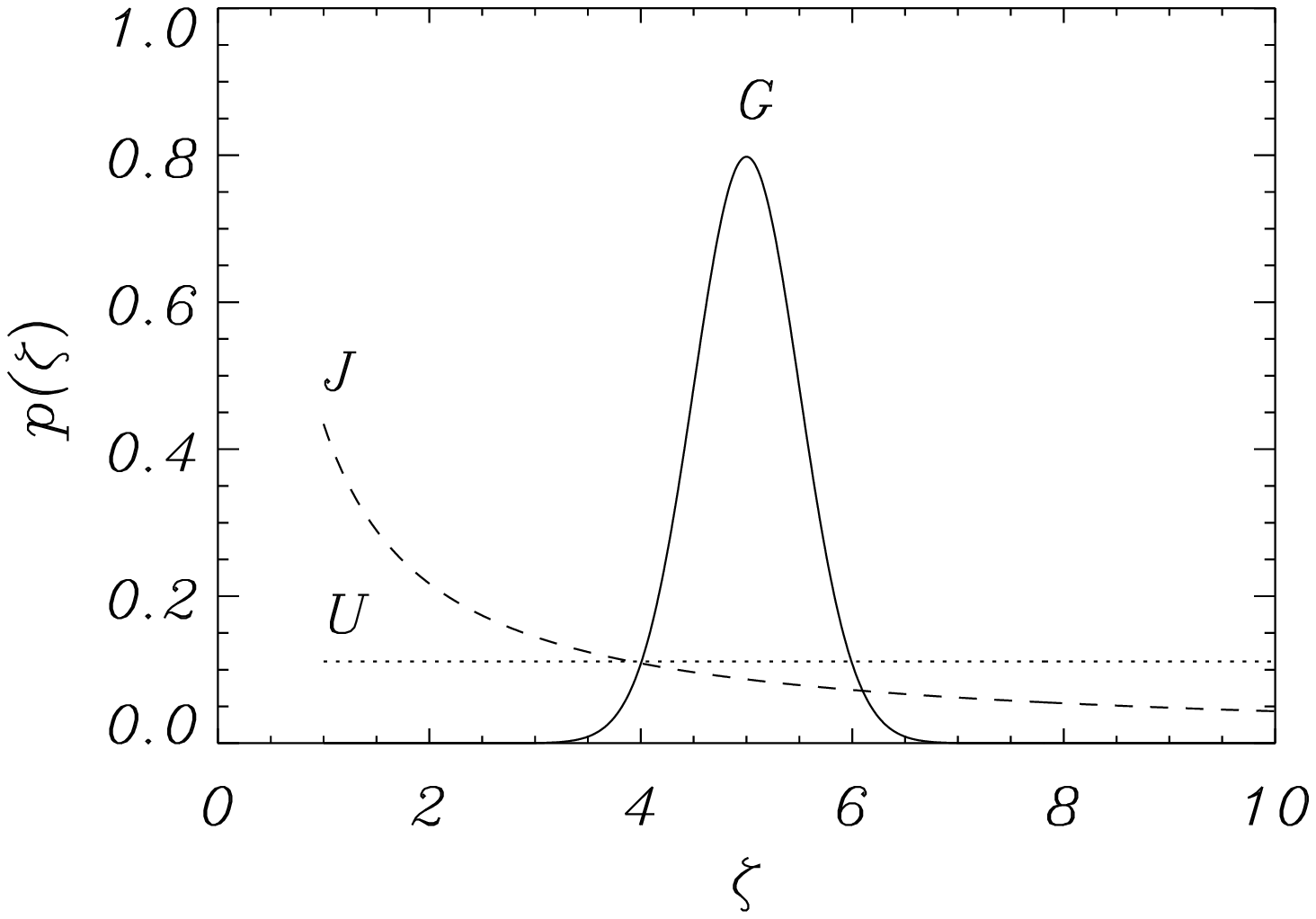} 
    \caption{Example of Uniform,  Jeffreys, and Gaussian prior probability distributions for the density contrast in the range $\zeta\in[1,10]$. For the Gaussian distribution, a mean value $\zeta=5$ with standard deviation $\sigma_\zeta=0.5$ has been considered.}
   \label{fig:priors}
\end{figure}

In our analysis,  we use a uniform prior for the transverse inhomogeneity length-scale and the internal Alfv\'en travel time. The limits for the transverse inhomogeneity are imposed by the modeling, since this parameter must be in the range $l/R\in[0,2]$. For the internal Alfv\'en travel time, we have considered uniform prior distributions with minimum and maximum values that enclose the intervals given by  $I_{\rm y}$ in expressions~(\ref{yzez1}), taking into account the observed period.  Coronal loops are over-dense structures with an internal density that is at most one order 
of magnitude larger that the external coronal density. Densities two orders of magnitude larger than the coronal density, typical in prominence plasmas, are less likely in coronal loops.
Our analysis makes use of the three prior distributions given by Equations~(\ref{uniform}), (\ref{jef}), and (\ref{gaussian}), for the density contrast. The virtues and disadvantages of each of them are discussed in Section~\ref{synthetic}.
Figure~\ref{fig:priors} displays an example of the three different types of prior in density contrast. In each case, the integral of the prior probability over the entire range of values must amount to one.

\subsection{Markov Chain Monte Carlo method}

The objective of our parameter inference under the Bayesian framework is to sample the full posterior distribution so that, for instance, we can
locate the most plausible model that maximizes Equation~(\ref{bayes}) or calculate the marginalization integrals. To this end, one has to evaluate  
$p$({\boldmath $\theta$}$|d$) for  different combinations of parameters. If ten values for parameter are to be  evaluated and N$_{\rm par}$ is 
the number of parameters this means that $\sim10^{N_{\rm par}}$ evaluations are needed. This exponential increase of the number of evaluations with 
the number of parameters is known as the curse of dimensionality. A convenient way to handle this kind of problem is to perform  a Markov Chain
Monte Carlo (MCMC) simulation,  for the sampling of the posterior probability distribution function. The MCMC algorithm we apply 
allows us to construct a sequence of points in parameter space, called a ``chain'', whose density is proportional to the posterior distribution function. 
It therefore provides a method for sampling the posterior distribution up to a multiplicative constant. We use the same  code employed by 
\cite{andres07} in their inversion of Stokes profiles. The details of the technique and the algorithm are given  in that paper, so only the most 
relevant information is detailed here. 

The obtained Markov chain is defined as a sequence of random variables  ({\boldmath $\theta_0$}, {\boldmath $\theta_1$}, $\ldots$, {\boldmath $\theta_n$}), 
such that the probability of the {\boldmath $\theta_i$} element in the chain only depends on the value of the previous element, 
{\boldmath $\theta_{i-1}$}. Starting from a given element in the chain, the next point is chosen in such a way that the distribution 
of the chain asymptotically tends to be equal to the posterior distribution. Markov chains can be shown to converge to a stationary 
state where successive elements of the chain are samples from the target distribution, in our case the posterior $p$({\boldmath $\theta$}$|d$).

Our method uses the Metropolis algorithm \citep{Metropolis53,Neal93}. Starting from a given vector of parameters {\boldmath $\theta_0$},
the posterior probability given the data is calculated, $p$({\boldmath $\theta_0$}$| d$). This includes the calculation of the priors and the likelihood, which involves the evaluation of the forward problem. Next, a new vector of parameters, {\boldmath $\theta_i$}, is obtained sampling from a proposal density 
distribution, $q$({\boldmath $\theta_i$}$|${\boldmath $\theta_{i-1}$}). Again the posterior is evaluated, $p$({\boldmath $\theta_i$}$| d$). The ratio

\begin{equation}
r=\frac{p( \mbox{{\boldmath $\theta_i$}}|d)q( \mbox{{\boldmath $\theta_i$}}|\mbox{{\boldmath $\theta_{i-1}$}})}{p( \mbox{{\boldmath $\theta_{i-1}$}}|d)q( \mbox{{\boldmath $\theta_{i-1}$}}|\mbox{{\boldmath $\theta_i$}})}
\end{equation}
is evaluated and {\boldmath $\theta_i$} is admitted with probability $\beta=\mathrm{min}[1,r]$. The process further progresses by obtaining a new vector of parameters from the proposal density distribution.

The key ingredient in this process is the proposal density $q$({\boldmath $\theta_i$}$|${\boldmath $\theta_{i-1}$}). This distribution is used to sample points from the posterior
starting from a given point. Ideally, it should be chosen as close as possible to the unknown posterior distribution. Also, it should be simple to sample from it. Since the first condition is impossible 
to fulfill unless the posterior is known, we choose a sufficiently general distribution function which is easy to sample from. The obvious selection is a multivariate Gaussian distribution. In our case, we
allow for a non-diagonal covariance matrix in the Gaussian distribution, which improves the sampling efficiency of the MCMC code \citep[see Appendix A in][for more details]{andres07}.  We have chosen a uniform distribution for the initial $N_{\rm unif}$ steps of the chain. Once some information about the posterior is known the algorithm changes to a Gaussian proposal density centered on the current value of the parameter.

\section{Statistical inversion results}\label{results}

\subsection{Synthetic data}\label{synthetic}

In order to assess the performance of our MCMC code, when applied to the problem at hand, the inversion of physical parameters is first performed with synthetic data for period and damping times. These synthetic data are obtained from the solution to the forward problem (expressions~[\ref{forward}]) with known values of the loop parameters, so that the numerical inversions are performed under controlled conditions. In addition, this approach enables us to test the performance of the three different prior probability distributions for the density contrast discussed in Section~\ref{priors}. 

We have considered a model coronal loop with the following physical parameters: $\zeta=5$, $l/R=0.25$, and $\tau_{\rm Ai}=150$ km s$^{-1}$. Forward modeling, according to the theory of resonantly damped kink waves under the thin tube and thin boundary approximations,  predicts a period  $P=232.4$ s and a damping ratio  $\tau_{\rm d}/P=3.8$. The period and the damping ratio are then used as data for the inversion and the ability of the code to recover the model coronal loop properties is analyzed, for different prior distributions and varying ranges for their definition. In all our simulations, variances of $\sigma_P=0.1P$ and $\sigma_{\tau_{\rm d}}=0.1\tau_{\rm d}$ are considered.

\begin{figure}[t]
   \centering
  \includegraphics[width=8cm,height=8cm]{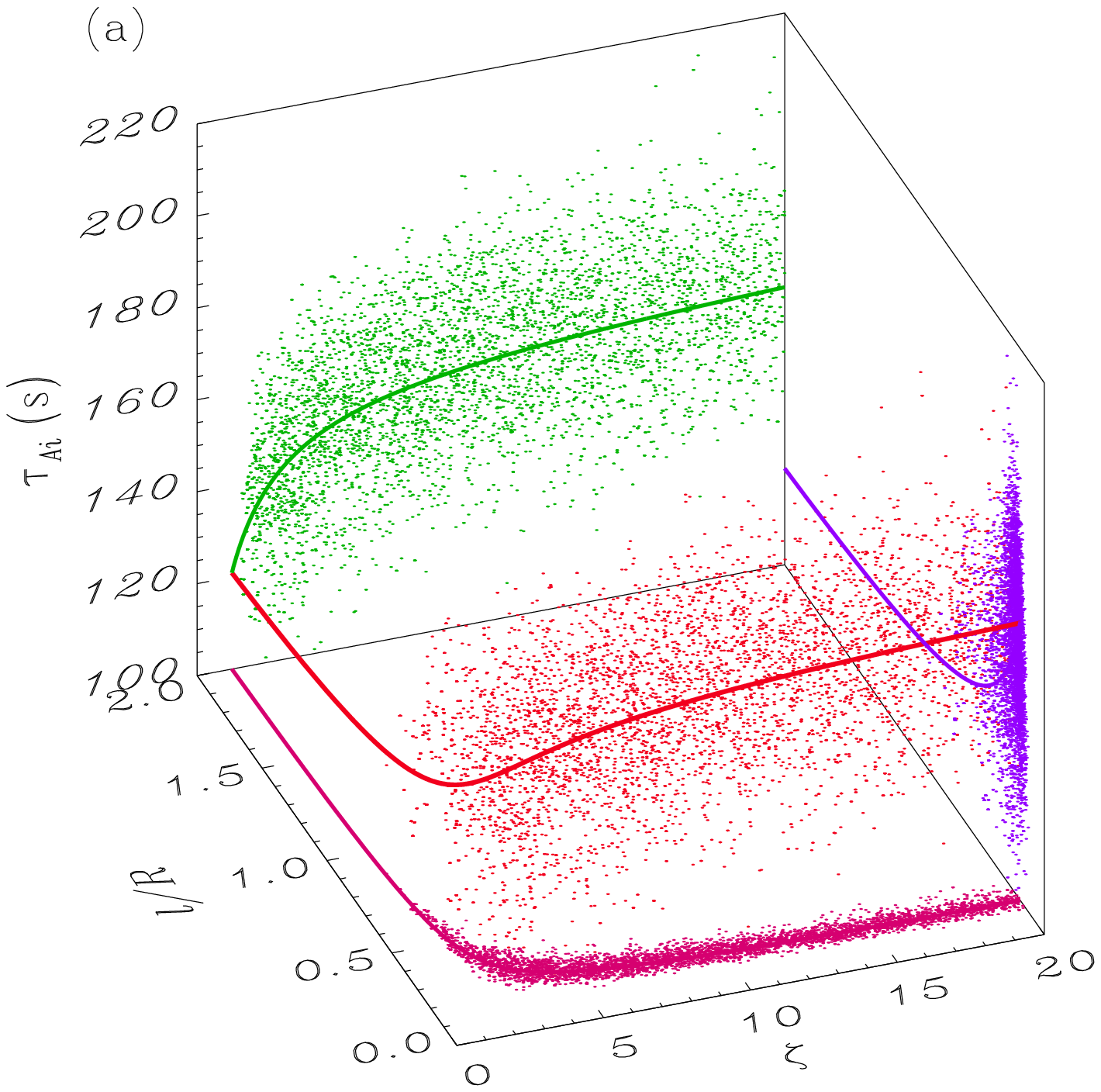} 
   \includegraphics[width=8cm,height=8cm]{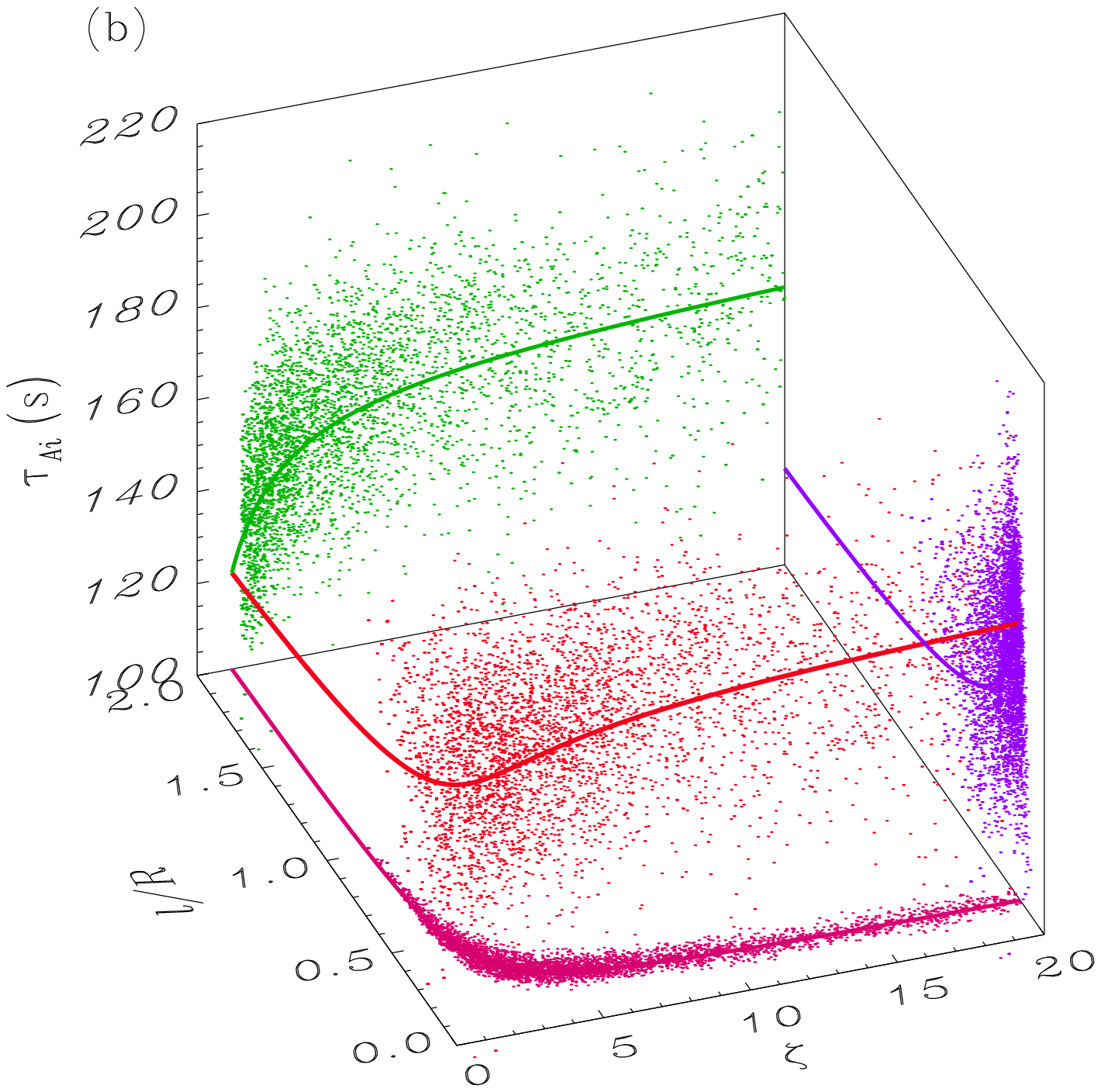} \\ 
   \includegraphics[width=8cm,height=8cm]{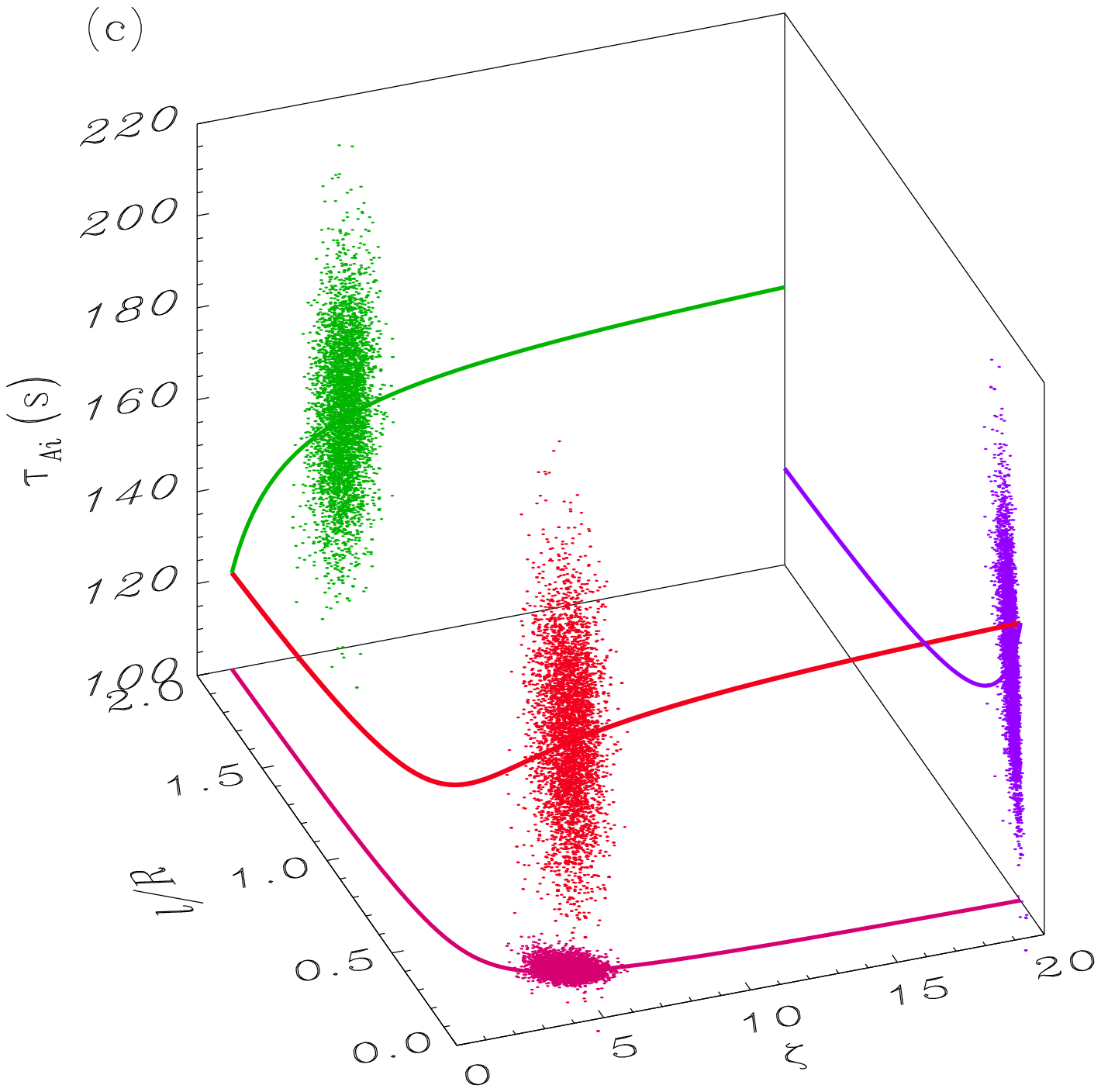} 
\caption{Three-dimensional view (and projections onto the different planes) of the converged Markov chains in the ($\zeta$, $l/R$, $\tau_{\rm Ai}$) parameter space 
                  for a synthetic coronal loop with $\zeta=5$, $l/R=0.25$, and $\tau_{\rm Ai}=150$ km s$^{-1}$. 
                  The  inversions were performed using uniform prior distributions for  $l/R\in[0,2]$ and $\tau_{\rm Ai}\in[1,400]$ and three different prior distributions  for $\zeta\in[1.2,20]$:
                  (a)  uniform prior; (b) Jeffreys prior; and (c) Gaussian prior centered at $\zeta=5$ and variance $\sigma_{\zeta}=0.1\zeta$. 
                  In all plots, solid-lines correspond to the analytic inversion curve and variances of $\sigma_P=0.1P$ and $\sigma_{\tau_{\rm d}}=0.1\tau_{\rm d}$ have been used for the oscillation period and damping time.}
   \label{fig:3dchain-priors}
\end{figure}

\begin{figure*}[t]
   \centering
   \includegraphics[width=8cm,height=6cm]{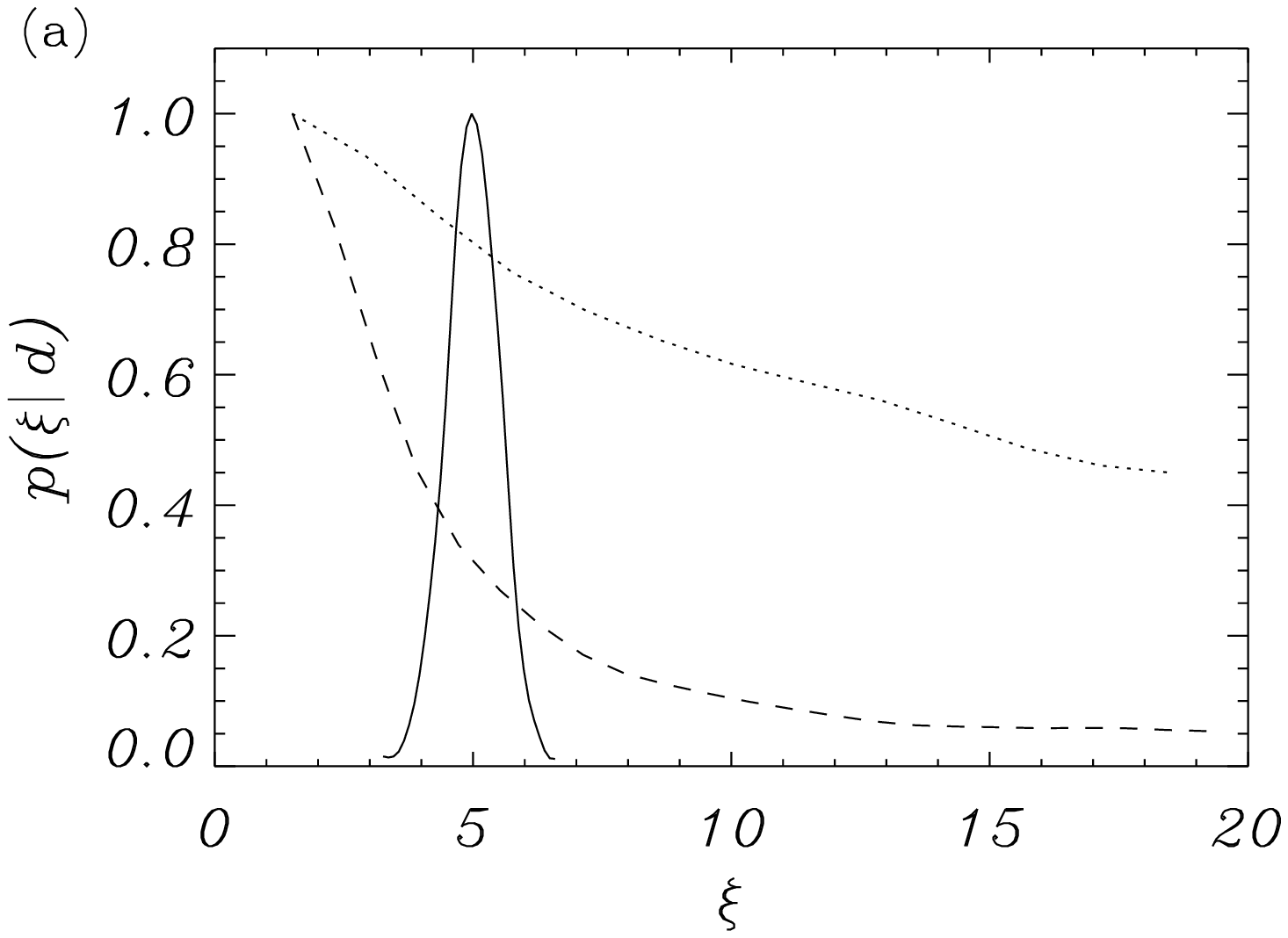} 
   \includegraphics[width=8cm,height=6cm]{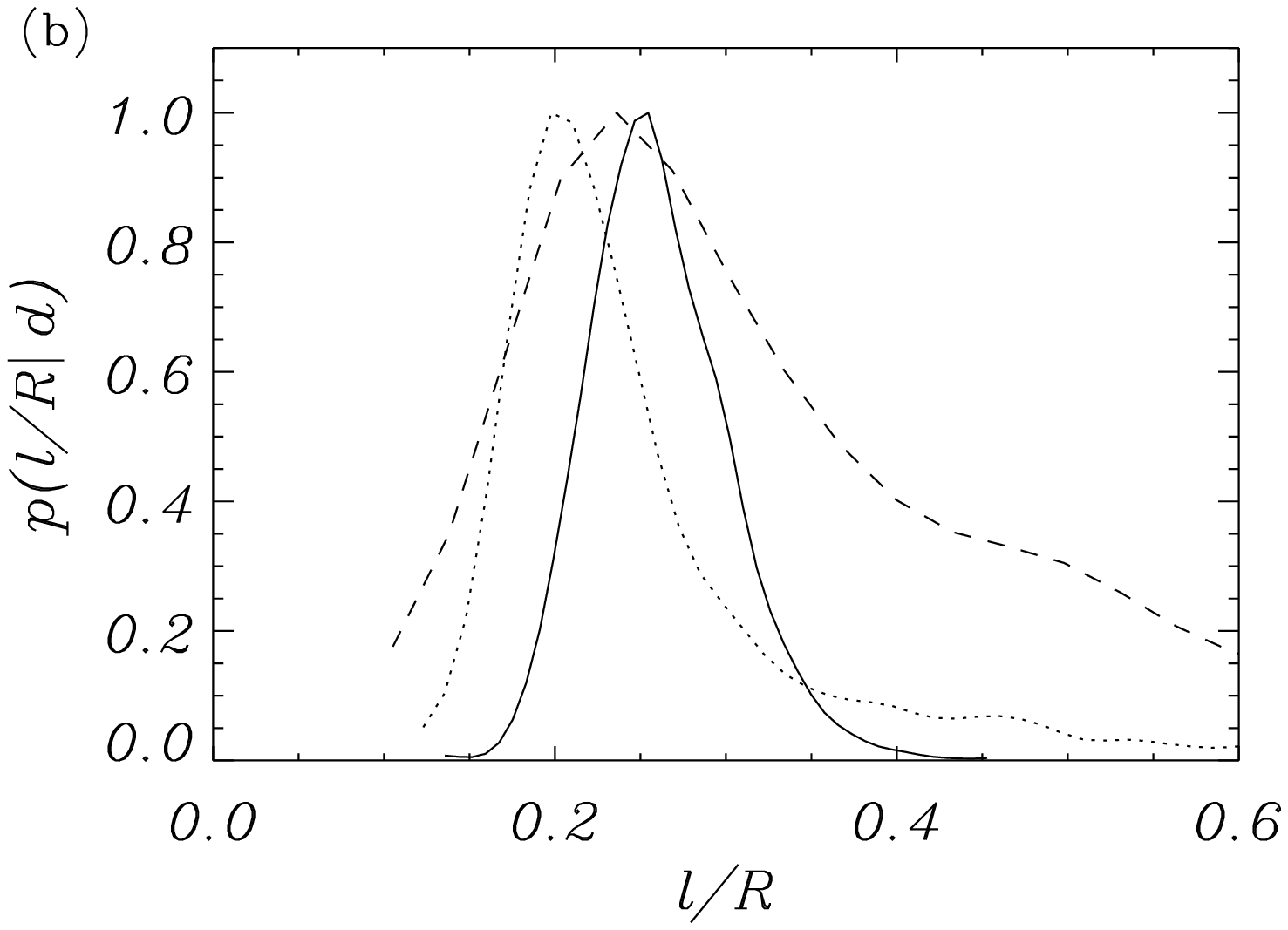} \\
   \includegraphics[width=8cm,height=6cm]{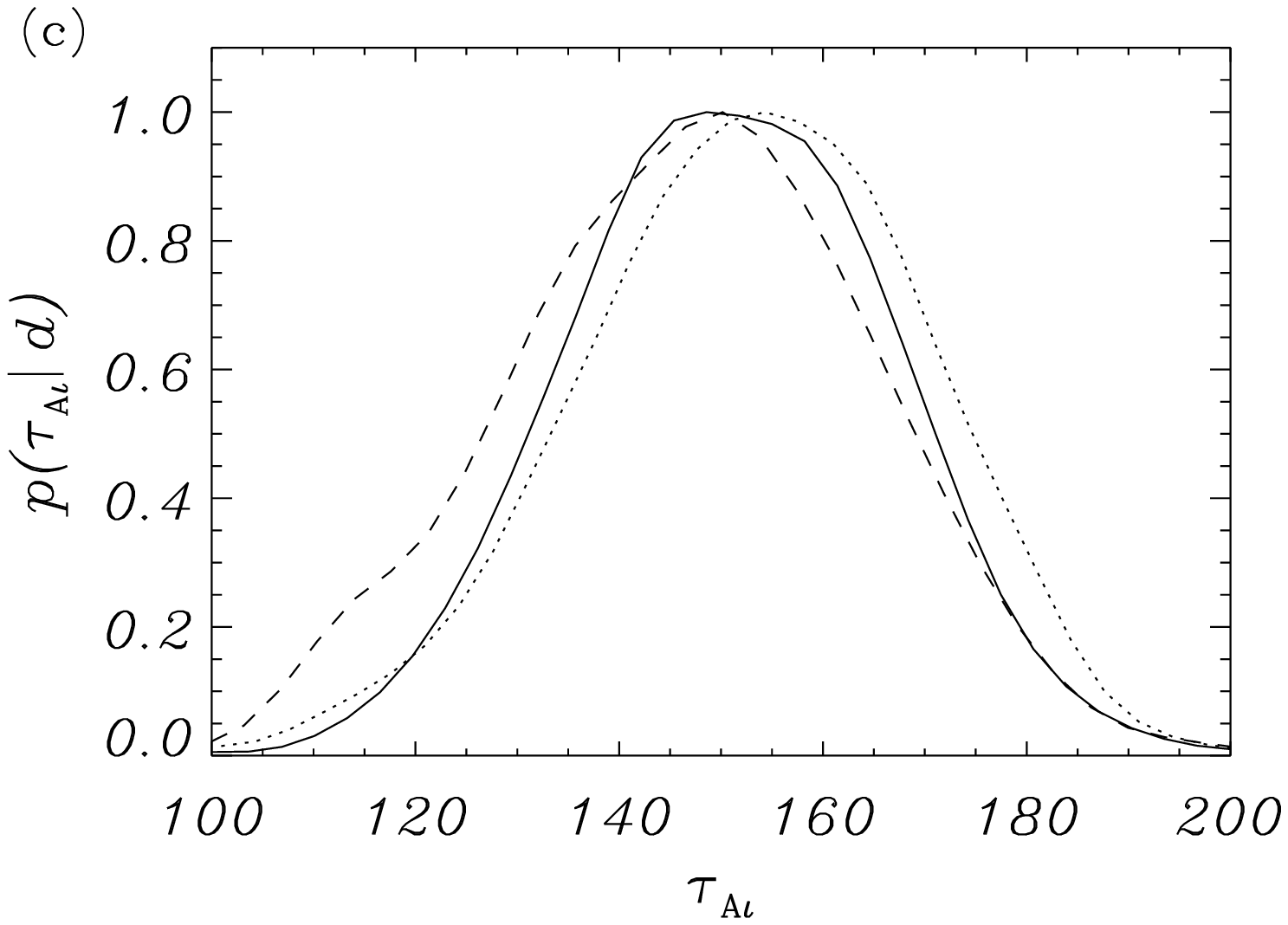} 

    \caption{One-dimensional marginalized posterior distributions for the density contrast (a), the transverse inhomogeneity length-scale (b), and the internal Alfv\'en travel time (c), corresponding to the inversions displayed in Figure~\ref{fig:3dchain-priors}. The solid, dashed, and dotted lines correspond to the three different prior probability distributions for density contrast, as indicated in Figure~\ref{fig:priors}.}
   \label{fig:inversion}
\end{figure*}

Figures~\ref{fig:3dchain-priors} and \ref{fig:inversion} show the inversion results obtained with these data, for fixed prior distributions for the transverse inhomogeneity length scale and internal Alfv\'en travel time. For the density contrast, the three different prior types defined in Section~\ref{priors} are used in the range $\zeta\in[1.5,20]$. In all three cases, the resulting Markov chains closely follow the same regions outlined by the analytic inversion curve.  The distribution and density of the elements in the chain - which directly determine the properties of the posterior - differ in different regions of the parameter space. For a uniform prior distribution in density contrast, the elements in the chain are roughly uniformly distributed along the inversion curve (Figure~\ref{fig:3dchain-priors}a). The relevant quantitative information from the inversion 
comes from the analysis of the posterior probability distribution functions. Since the elements of the Markov chains are samples from the full posterior it is easy to divide the range for a given parameter in a series of bins and count the number of samples falling within each bin. By computing the marginal posterior for each parameter, we see that the density contrast cannot be constrained (see Figure~\ref{fig:inversion}a). However, the projection of the chain onto the ($l/R$, $\tau_{\rm Ai}$)- plane indicates that a well defined posterior probability distribution can be obtained for these two parameters. Figures~\ref{fig:inversion}b and c show that, indeed, a very good estimation of these
two parameters is possible. When a Jeffreys type prior for the density contrast is used, the elements in the chain are not uniformly distributed. Their density appears to be larger at positions in the parameter space that are close to the model coronal loop properties (Figure~\ref{fig:3dchain-priors}b).  The marginal posterior for the density contrast, however, does not show a well defined probability distribution (Figure~\ref{fig:inversion}a). The remaining two parameters, though, are  well constrained (Figures~\ref{fig:inversion}b and c). Finally, we introduce a hypothetical density contrast measurement for our model coronal loop and use a Gaussian prior distribution for the density contrast centered around that measurement. Figure~\ref{fig:3dchain-priors}c shows that in that case, the inversion produces a Markov chain whose elements are closely packed around the correct loop properties. The marginal posteriors in this case (Figure~\ref{fig:inversion}) show well-defined Gaussian-like distributions for the three parameters of interest.

These results indicate that, in general, the Bayesian inversion scheme will enable us to constrain two out of the three parameters of interest, the transverse inhomogeneity length-scale and the internal Alfv\'en travel time. The density contrast, in general, cannot be constrained. If additional information about the internal and external densities is available from observations, this information can then be used to fully constrain the three unknowns. From the obtained posterior distributions in Figure~\ref{fig:inversion}, the median and the computation of the 68\% confidence level can be used to obtain estimates with error bars for the inferred parameters.

\begin{figure*}[t]
   \centering
   \includegraphics[width=5cm,height=4cm]{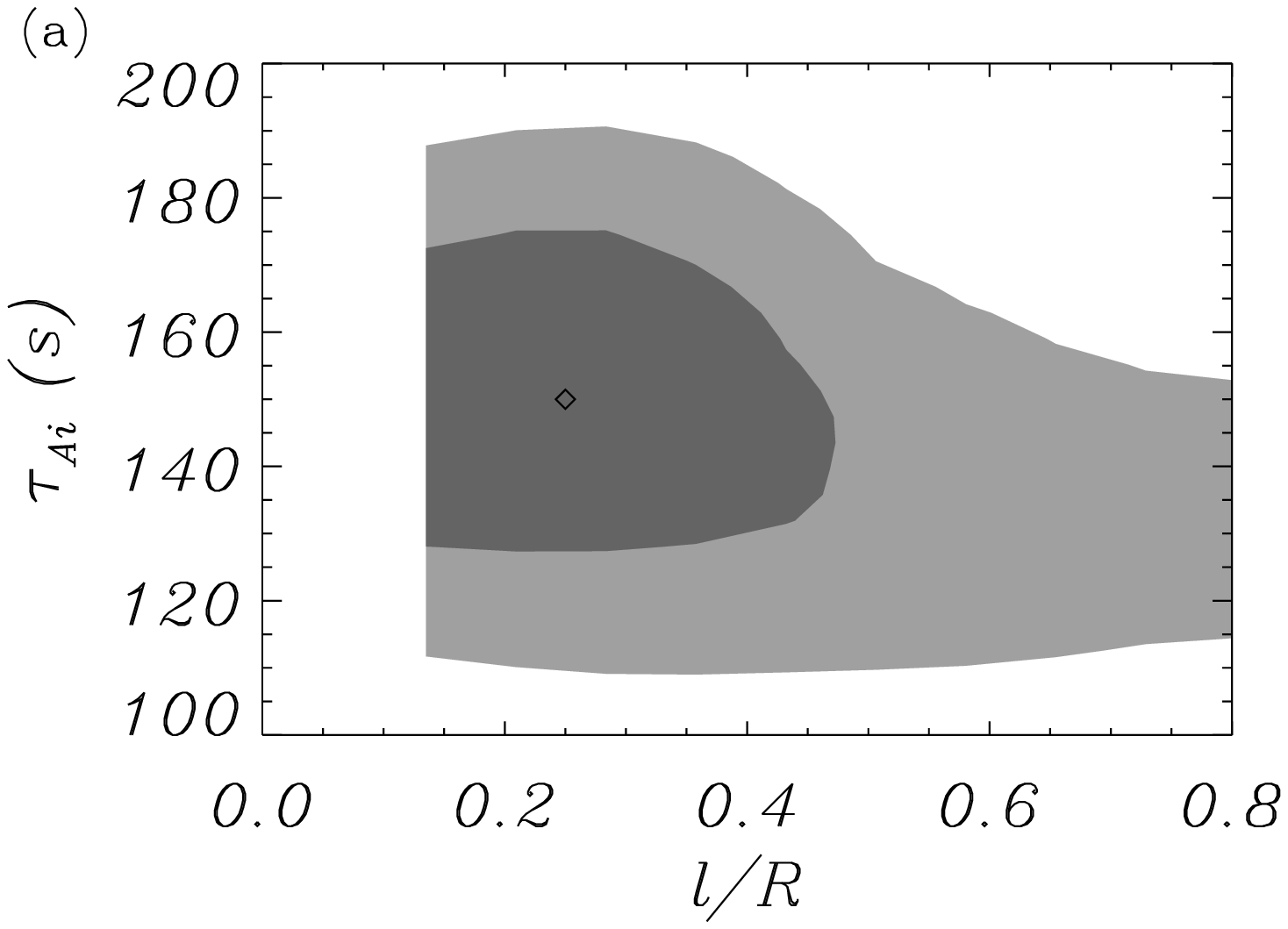} 
   \includegraphics[width=5cm,height=4cm]{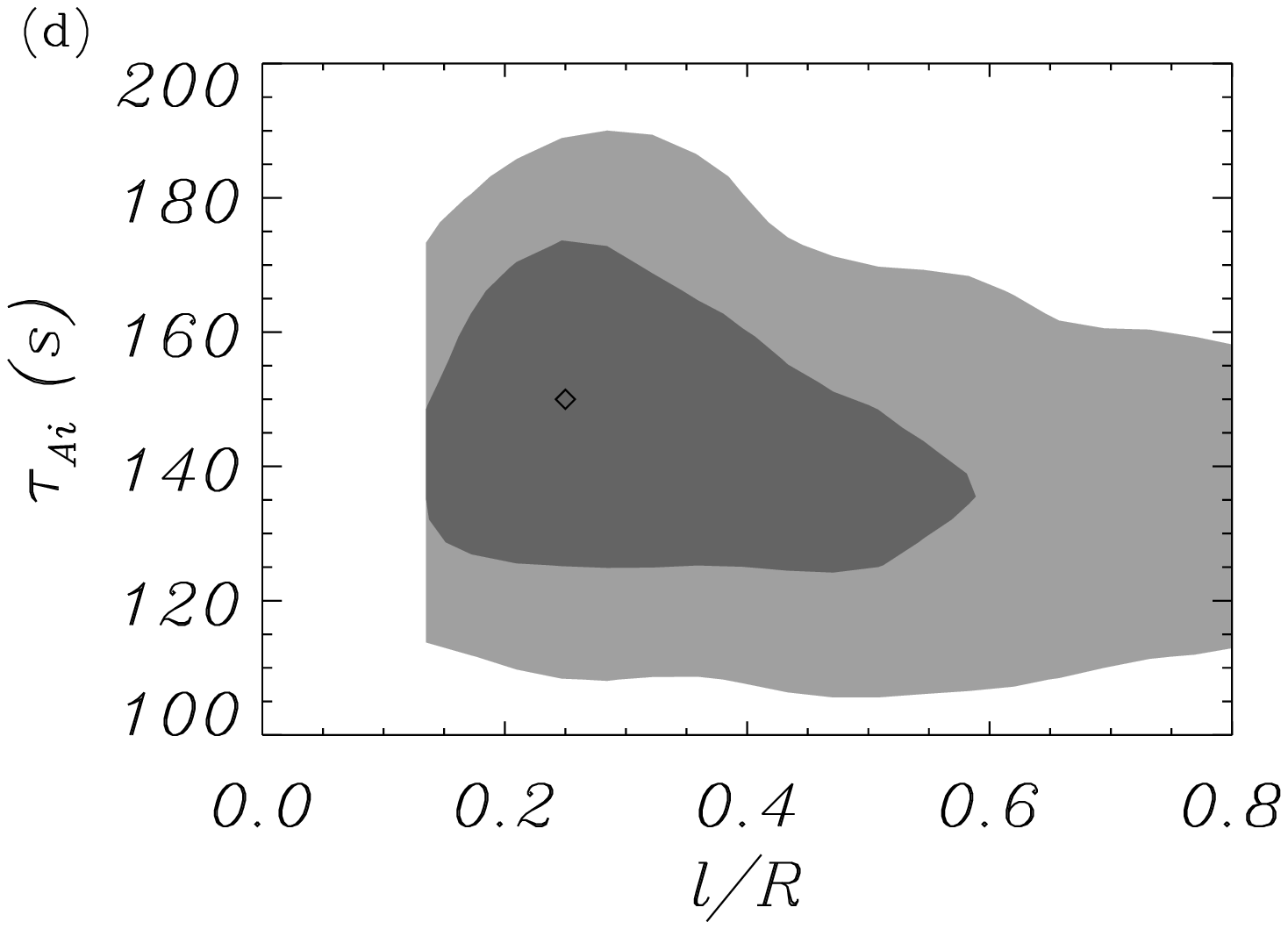} 
   \includegraphics[width=5cm,height=4cm]{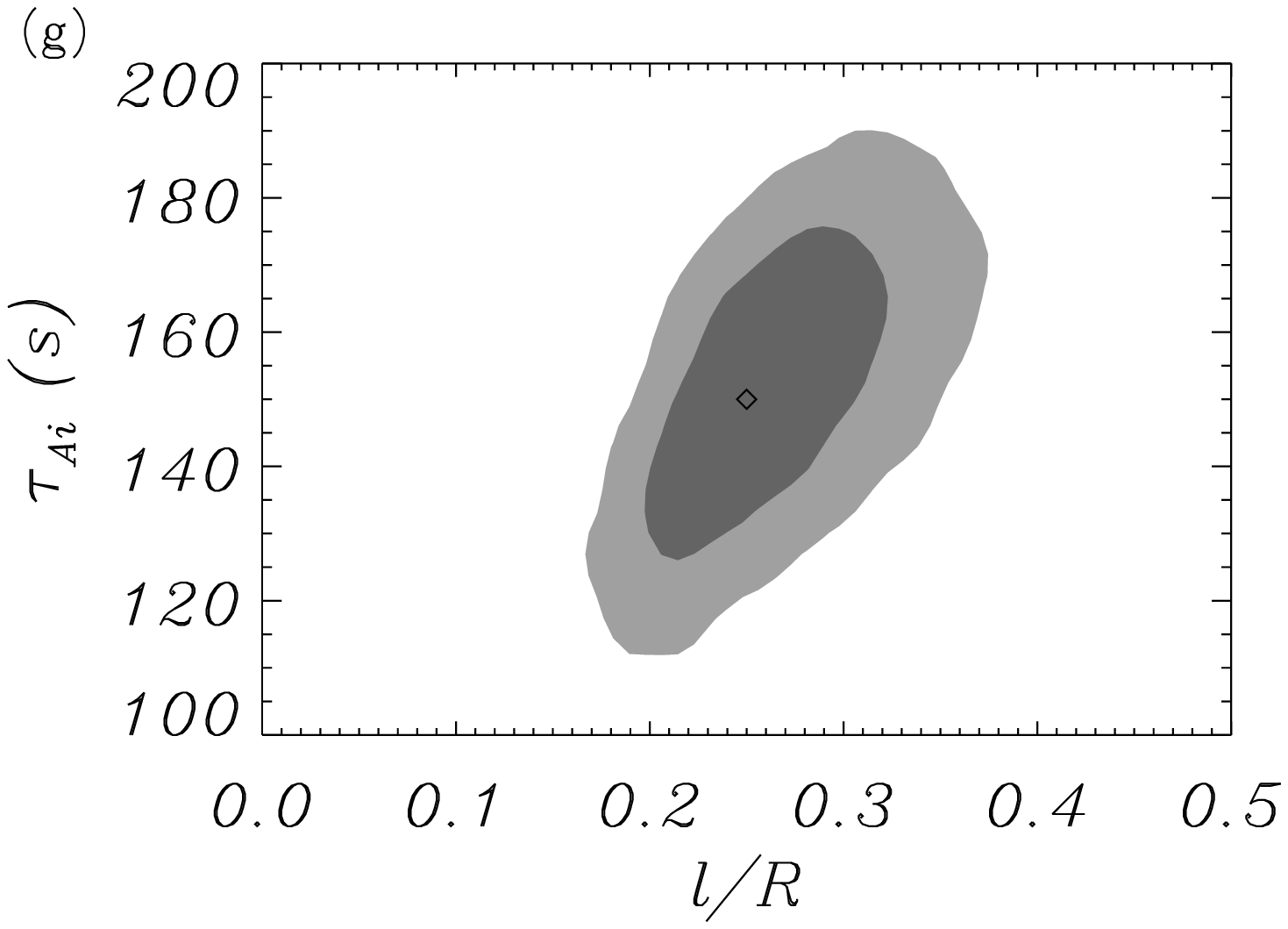} \\
    \includegraphics[width=5cm,height=4cm]{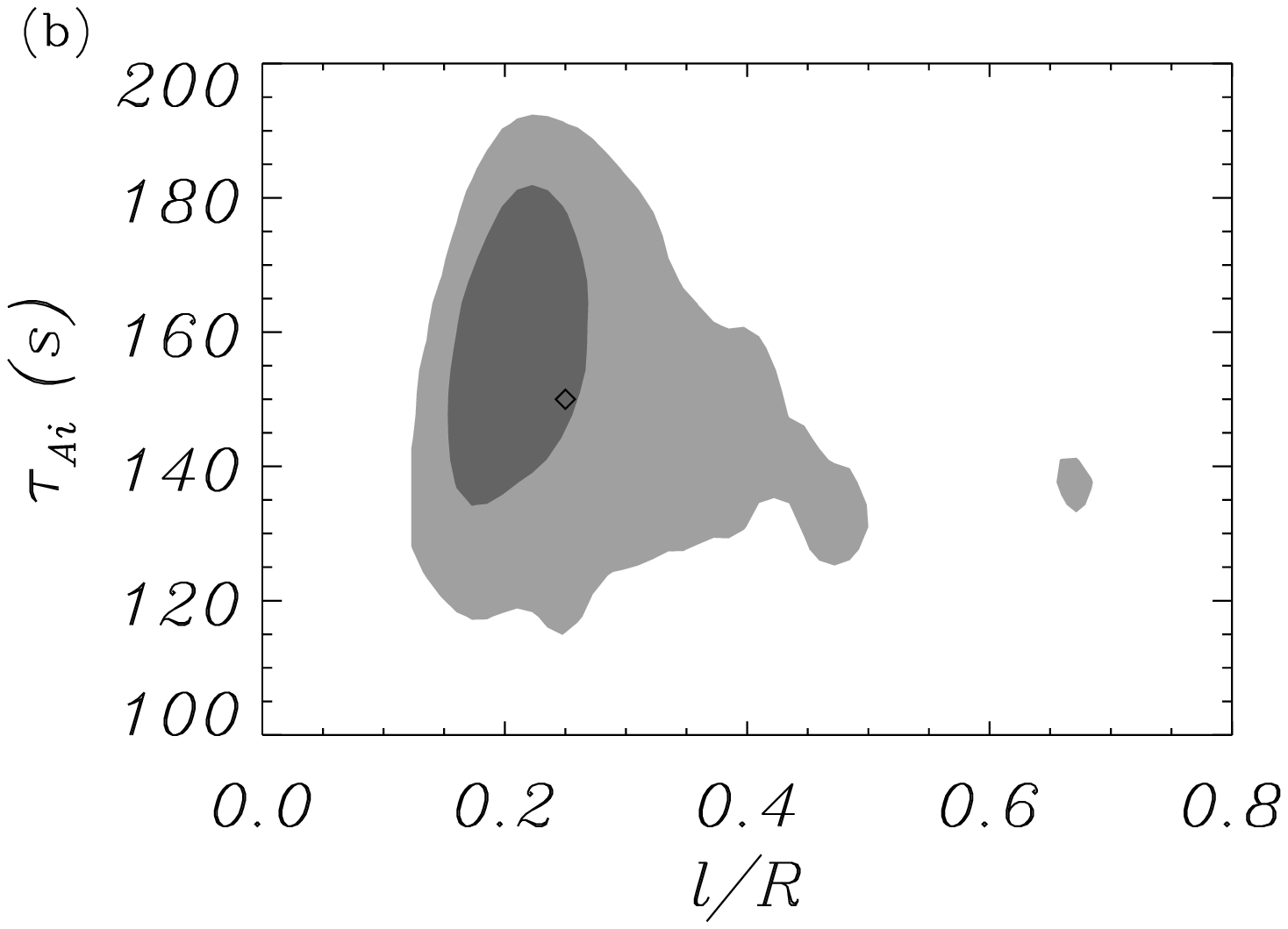} 
   \includegraphics[width=5cm,height=4cm]{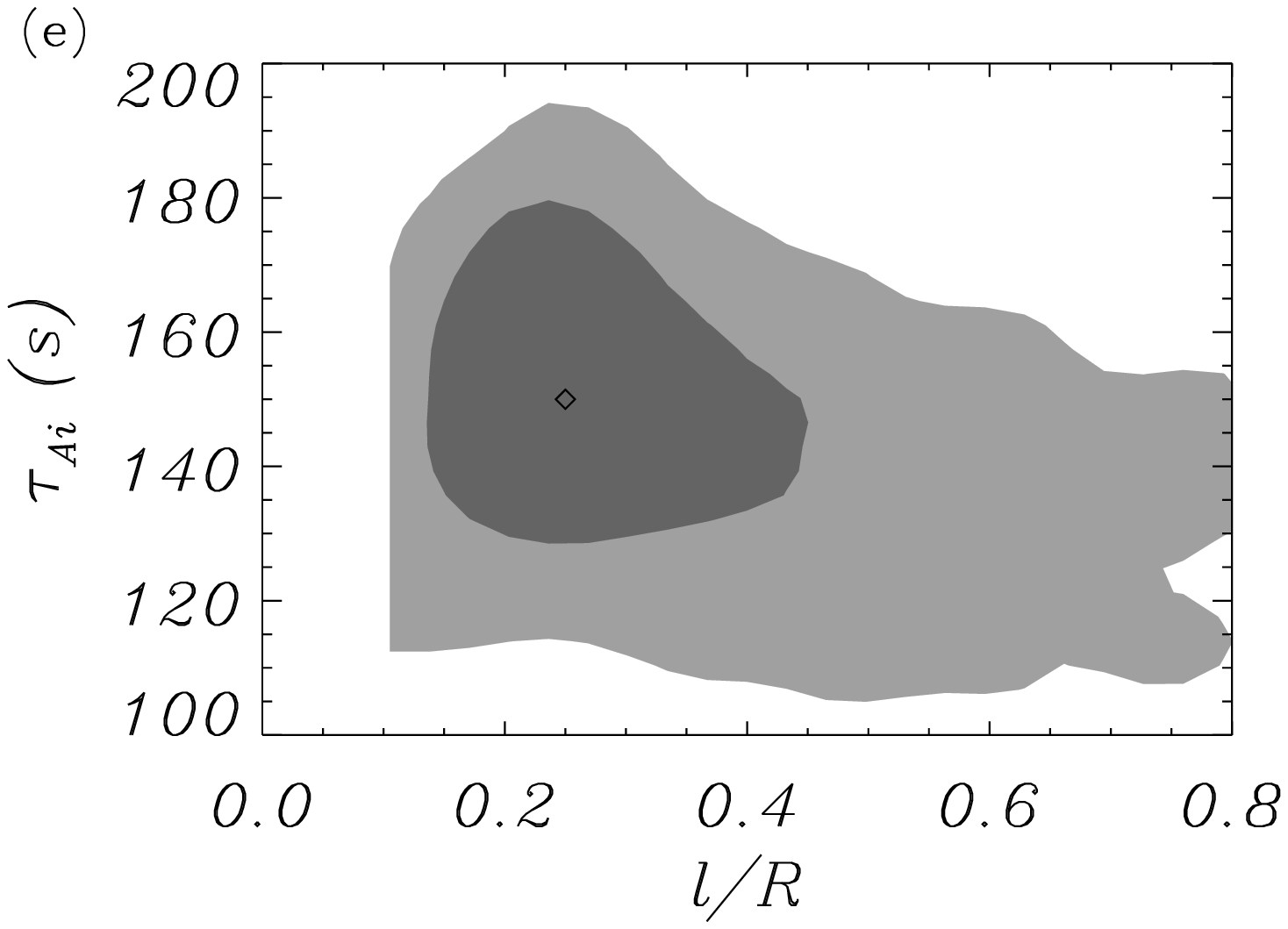} 
   \includegraphics[width=5cm,height=4cm]{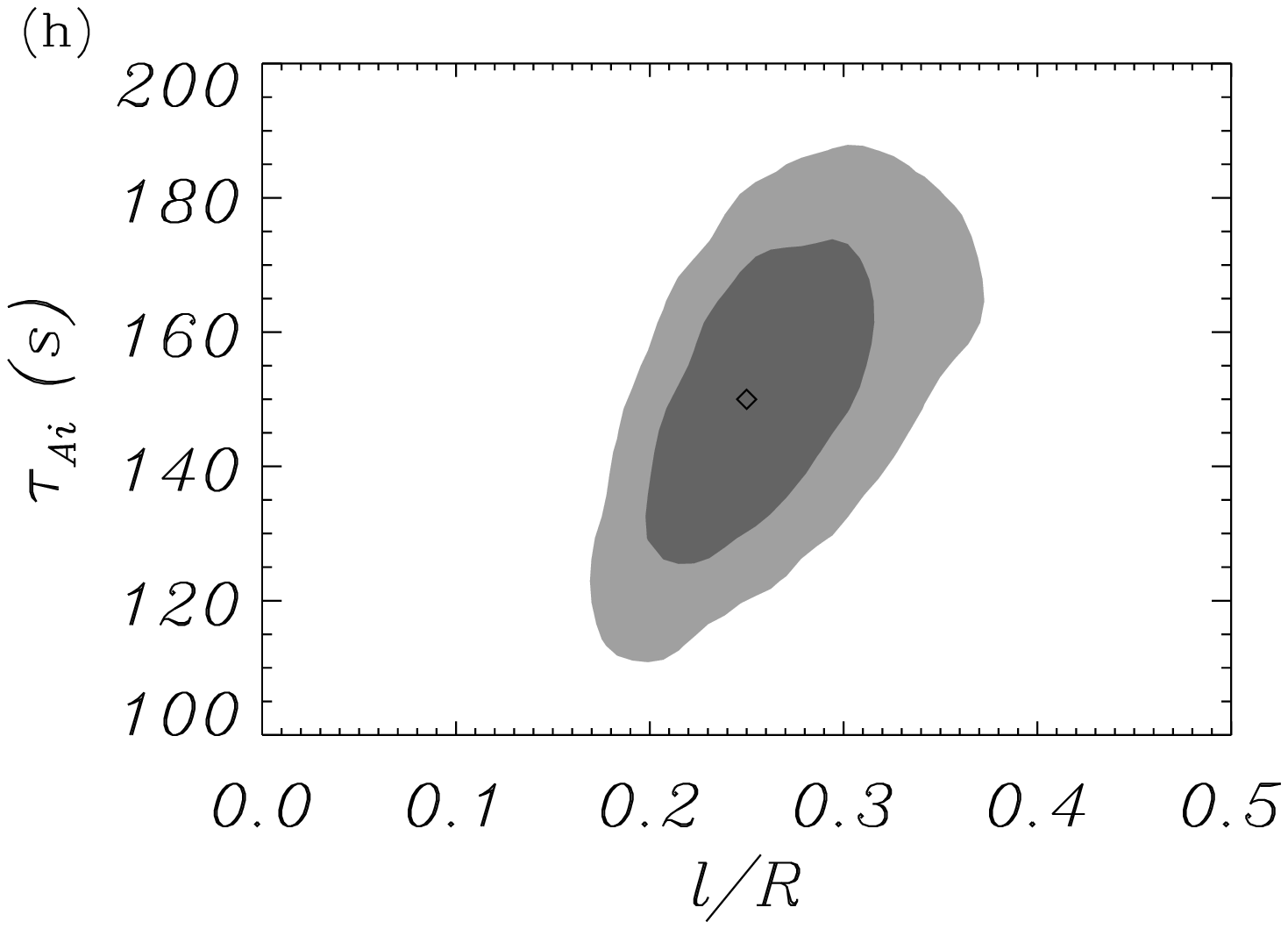} \\ 
   \includegraphics[width=5cm,height=4cm]{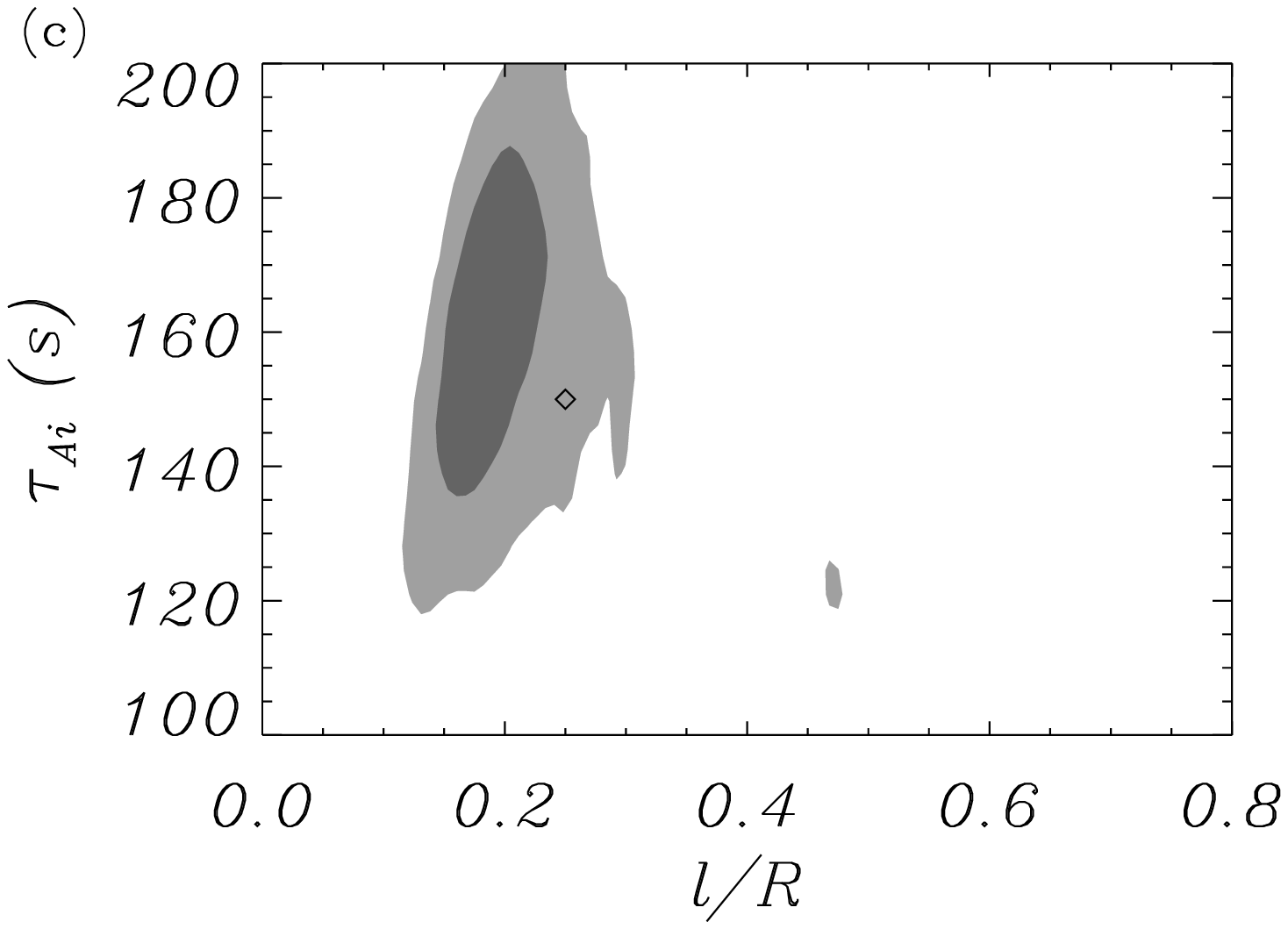} 
   \includegraphics[width=5cm,height=4cm]{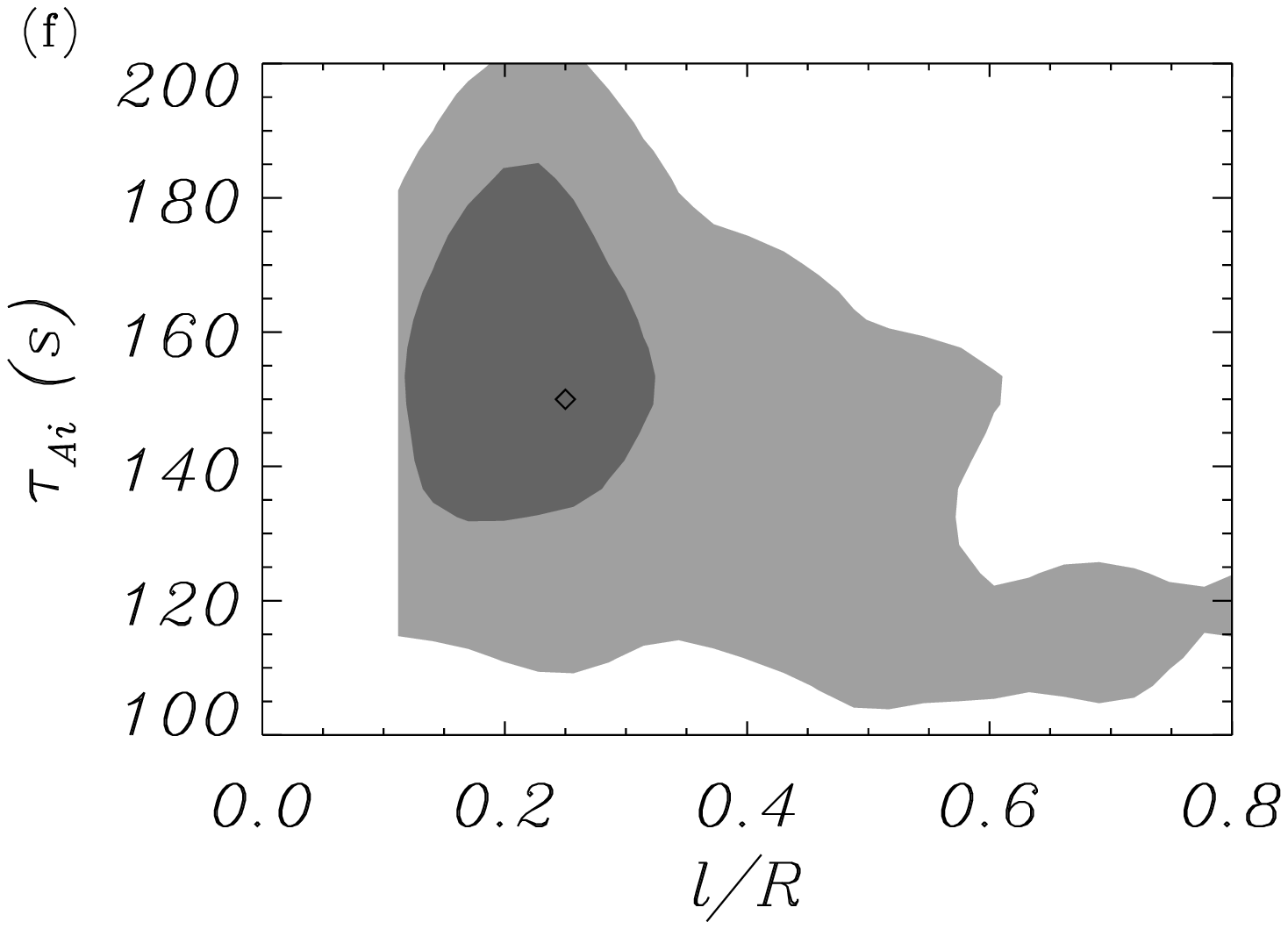} 
   \includegraphics[width=5cm,height=4cm]{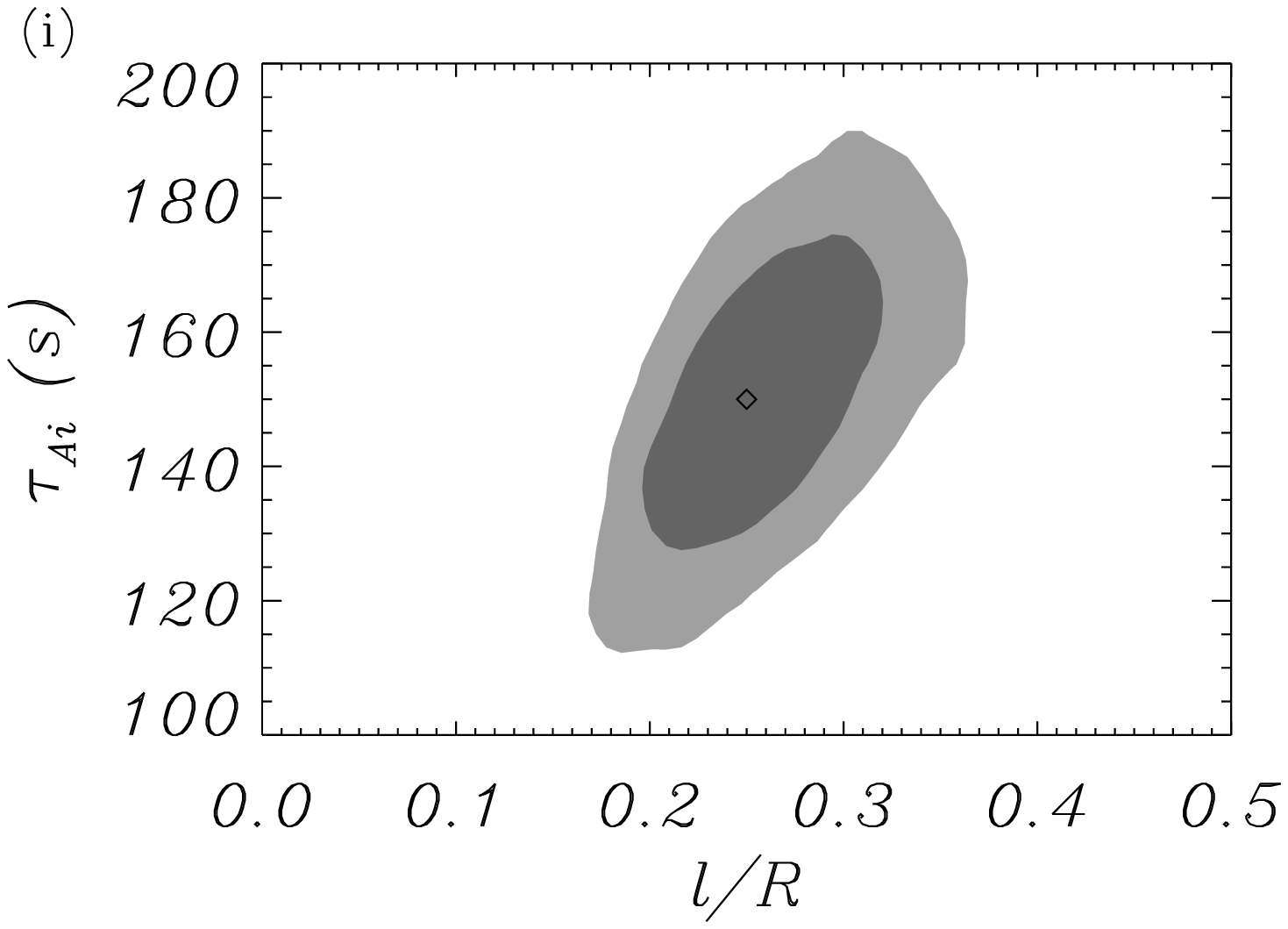} \\

    \caption{Joint two-dimensional posterior distributions for the transverse inhomogeneity length-scale and the internal Alfv\'en travel time for different combinations of prior types and ranges of variation for the density contrast. From left to right, (a)-(c) uniform prior;  (d)-(f) Jeffreys prior; and  (g)-(i) Gaussian prior with $\mu_\zeta=5$ and $\sigma_\zeta=0.5$ . From top to bottom,  (a)-(g) $\zeta\in[1.2,10]$;  (b)-(h) $\zeta\in[1.2,20]$; and  (c)-(i) $\zeta\in[1.2,50]$. The outer boundaries of the light grey and dark grey shaded regions indicate the 95$\%$  and 68$\%$ confidence levels. Symbols indicate the synthetic coronal loop properties, $l/R=0.25$, and $\tau_{\rm Ai}=150$ km s$^{-1}$.}
   \label{fig:joints}
\end{figure*}

We found that the obtained results are largely insensitive to the variation of the range assumed for the internal Alfv\'en travel time, as long as this range is sufficiently wide so it encloses the tails of the resulting posterior distribution. As for the density contrast, additional inversions were performed for different values of the upper limit in the prior distribution for this parameter, considering the three prior types. The results indicate that some prior types perform better than others. Figure~\ref{fig:joints} displays the joint probability distribution for the internal Alfv\'en travel time and the transverse inhomogeneity in the form of contours for the 68\% and 95\% confidence levels, for some illustrative cases. In all plots the known coronal loop parameters are indicated by a symbol. Numerical values for all performed inversions are given in Table~\ref{table:synthetic}.

When a uniform prior for density contrast is considered, increasing the upper limit of the considered range (Figures~\ref{fig:joints}a-c) produces a displacement of the joint marginal posterior, in such a way that the inversion underestimates the transverse inhomogeneity length scale and overestimates the Alfv\'en travel time. Although the real values are for most of the cases enclosed within the error bars, we can see that for example, for the largest considered range (Figure~\ref{fig:joints}c), the real value is already outside the 68\% confidence level given by the inversion. Another effect that can be appreciated looking at Figures~\ref{fig:joints}a-c is the shrinking of the joint posterior for $l/R$, that results in smaller error bars for this parameter (see also Table~\ref{table:synthetic}). This result cannot be taken as
an improved inversion, as is solely due to the fact that by extending the range in $\zeta$, the uniform prior  gives more weight to elements in the chain with large density contrast, hence 
lower transverse inhomogeneity length scales and larger Alfv\'en travel times are obtained. As a matter of fact, the integral of the prior distribution in Equation~(\ref{uniform}) in the range $\zeta\in[10,100]$ is about ten times the same integral in the range $\zeta\in[1,10]$. The optimal results for the inversion using a uniform prior in density contrast are obtained when the upper limit in $\zeta$ is around 2 or 3 times the real density contrast value. The problem is that in a real application we can hardly know the real density contrast value.

A better performance and independence of the results on the prior information is obtained when considering a Jeffreys type prior for density contrast. Since this distribution assigns decreasing probability for increasing contrast, the integrals of the prior in Equation~(\ref{jef}) in the ranges $\zeta\in[1,10]$ and $\zeta\in[10,100]$ are of the same order. Figures~\ref{fig:joints}d-f show the results for this prior. It can be appreciated that the obtained marginal posteriors are almost independent of the assumed range on the prior information, 
once the upper limit for $\zeta$ is sufficiently large, and that optimal results that recover the known coronal loop parameters are obtained (Table~\ref{table:synthetic}). For this reason, even if the density contrast itself cannot be constrained, the use of a Jeffreys prior in density contrast is found to be appropriate to perform the inversion of the remaining two parameters in real coronal loops, in the case information about their density is lacking or uncertain.

When density measurements are available, the Gaussian prior defined in Equation~(\ref{gaussian}) can be used. We have tested the performance of the inversion using this type of prior. As Figures~\ref{fig:inversion}, \ref{fig:joints}g-i, and Table~\ref{table:synthetic} show, in this case a full determination of the three unknowns is obtained, that remarkably well recovers the known coronal loop parameters. In addition, the inversion results are independent on the underlying assumptions of the a priori ranges in density contrast. Caution is called to the fact that, in this case, the reliability of the inferred parameters is entirely dependent on the reliability of the density contrast measurement.

Additional inversions under controlled conditions were performed for different combinations for the model coronal loop properties with, e.g.,   $\zeta=2.5, 10, 15$ and $l/R=0.5, 0.85$.
The obtained results do not change in a qualitative way those described above.

\subsection{Coronal loop oscillations}

Once the performance of our inversion method is evaluated, using synthetic data, we have applied the Bayesian inversion scheme to the 11 loop oscillation events  analyzed 
in e.g., \cite{OA02,GAA02,Aschwanden03b,Arregui07, GABW08}. Table~\ref{restable} displays the main oscillation properties for these events and the results from the application of analytic and Bayesian inversion techniques. Estimated values correspond to the median of the obtained distribution and errors are given at 68\% confidence level.

Observed period and damping ratios are used as data in our inversion code. Measurement errors in observed periods and damping times for oscillating coronal loops are lacking in the literature. \citet{Aschwanden02} presented the most complete catalog of events with measurements of periods and damping times for 26 oscillating coronal loops. Uncertainties up to 40-60\% can be found in those events for the measured periods and damping times. The analysis by \cite{vd07} drastically reduces the uncertainties to about 1\% in period and 3\% in damping time. It is not clear the meaning of such small errors on periods and damping times, with uncertainties that are shorter than the exposure time. In our analysis, we consider Gaussian errors of 10\% on both quantities.
Uniform prior distributions are considered for the transverse inhomogeneity length-scale, in the range $l/R\in[0,2]$, and for the internal Alfv\'en travel time, 
in a range determined by the oscillation period. We found that the ranges $\tau_{\rm Ai}\in[1,400]$ and $\tau_{\rm Ai}\in[1,800]$ easily accommodate the resulting 
posterior distributions for the shorter and longer period oscillation events in Table~\ref{restable}, respectively. As for the prior information in density contrast, based on 
our assessment in Section~\ref{synthetic}, we have performed the inversions using both a Jeffreys prior and a Gaussian prior, centered on the measurements 
reported by \cite{Aschwanden03b}, for the very same 11 loop oscillation events. For the Jeffreys prior the minimum, $\zeta_{\rm min}$, is computed using the
interval for $\zeta$ in expressions~(\ref{yzez1}). The maximum value is set to  $\zeta_{\rm max}=50$, based on our assessment with synthetic results (Table~\ref{table:synthetic}).
For the Gaussian prior, besides the estimated values we also use the computed error bars (see column ``Density ratio'' in Table 3 by \citealt{Aschwanden03b}), to implement the standard deviation in Equation~(\ref{gaussian}) for each case.  Uncertainties in density contrast listed by \cite{Aschwanden03b} can be as high as 50\%,  because of the indirect methods followed to obtain those estimates.
As the inversions using Gaussian prior information in density contrast are independent of the 
assumed ranges, we set the same range as before, $\zeta\in[\zeta_{\rm min},50]$. The use of a uniform prior in density contrast is discarded, because of its relatively 
poor performance in front of the Jeffreys prior, in the absence of observational information on this parameter.

Table~\ref{restable} first lists the analytically obtained intervals for internal Alfv\'en travel time computed using the corresponding expression 
for $y=\tau_{\rm Ai}/P$ in expressions~(\ref{yzez1}). 
This is all the information we can get from that inversion scheme, unless some additional assumption is made. Estimates for the Alfv\'en travel time using both Bayesian inversions, with Jeffreys and Gaussian priors in density contrast, very well accommodate within the analytically obtained intervals. These estimates are directly linked to the oscillation period, but in contrast to \cite{Nakariakov01}, the inversion now makes use of the information contained in both the period and the damping time. In addition, the Bayesian inversion enables us to compute error bars for this parameter, which are consistently calculated from the uncertainties on data. Error bars in the determination of the Alfv\'en travel time are found to be rather symmetric. 

We next consider the determination of the transverse inhomogeneity  length-scale. Regardless of the prior type used in the inversion, our estimates for $l/R$ are closely 
related to the observed damping rate, $P/\tau_{\rm d}$. Note that the larger this quantity, the stronger the damping by resonant absorption is and the larger the 
obtained inhomogeneity length-scales are. Error bars in the determination of $l/R$ are found to be rather asymmetric, with larger upper errors.  The 
reason is that  the posterior for $l/R$ has a more elongated tail towards the right of the maximum, because of the contribution of samples with low values of the density contrast (see the projection onto the [$\zeta$, $l/R$]-plane in Figure~\ref{fig:3dchain-priors}b).  The transverse inhomogeneity length-scale is a quantity that is meant to capture the radial variation of the local Alfv\'en frequency, which is unknown. Our modeling assumes a sinusoidal density variation. This being an assumption, it is not surprising that estimates for $l/R$ are subject to uncertainties. By comparing the estimates of $l/R$ obtained with Jeffreys and Gaussian priors, we find that  the numerical estimate considerably differs in some cases, e.g., for loops \#4, \#5, \#9, \#10, \#11. When the errors bars are taken into account they are rather compatible. More importantly, the concordance between damping ratio and inferred transverse inhomogeneity length-scale holds in both cases, in such a way that an ordering of inhomogeneities can be established by just following an ordering of damping ratios.

For the analyzed loop oscillation events, \cite{Aschwanden03b} additionally provide estimates for the transverse inhomogeneity length-scale, since the radius ($R$)  and the skin depth ($l$)
are measured (see their Table 2). These measurements lead to values of  $l/R$ in between $0.75$  and $0.96$, which in view of the values displayed on our Table~\ref{restable} would point to rather more transversely inhomogeneous loops, than those obtained from our inversions.  Notice that \cite{Aschwanden03b} have typical uncertainties of 20\% (up to 50\% in some cases) on their estimates of $l$.

Density contrasts in the last column  basically recover the input mean and standard deviation taken from the data measured by \cite{Aschwanden03b} and incorporated in the Gaussian prior. Their value is in the fact that they enable us to collapse the inversion chain, as explained in Section~\ref{synthetic}, and more accurately determine the remaining two parameters, thus producing the smaller error bars in $l/R$, in comparison with the Jeffreys prior results. As mentioned above, the reliability of the Gaussian inversion, especially in the determination of the transverse inhomogeneity length-scale, is entirely dependent on the reliability of the density contrast measurement. For this reason, it is reassuring that estimates of $l/R$ using both prior types - the less informative Jeffreys prior and the more informative Gaussian prior - result in similar values for most of the cases. On the other hand, because density measurements in the solar corona are challenging, we might consider that the inversion results using the less informative Jeffreys prior are more reliable that the ones that make use of uncertain 
information on density contrast. If that is the case, the differences in the determination of $l/R$ when comparing both Bayesian inversions could be ascribed to inaccurate density 
contrast measurement.

\section{Discussion and conclusions}\label{conclusions}

We have applied a Bayesian parameter inversion technique to the determination of unknown physical parameters 
in transversely oscillating coronal loops.  The model considers  coronal loops as one-dimensional cylindrically 
symmetric density enhancements with a radial density variation. The forward problem reduces to an algebraic 
set of equations for two observables as a function of three parameters. A Markov Chain Monte Carlo algorithm is used to sample 
the posterior probability distribution for the density contrast, the transverse inhomogeneity length-scale, and the internal Alfv\'en 
travel time. We find that the latter two parameters can be very well determined, while the density contrast cannot in general be constrained. 

Density contrast estimations are challenging in both observations and MHD seismology inversions.  By using synthetic data to perform the inversion under controlled conditions, different prior information for the density contrast has been tested. A uniform prior distribution is unable to correctly infer the parameters of interest, unless
some a priori information on density contrast is known, since it produces biased results  when extended ranges are considered. On the other hand, a Jeffreys type prior is more appropriate to perform the inversion when information on densities is lacking or uncertain.  If additional information on density contrast is available from observations, the three unknowns can be very well determined, by using a Gaussian prior centered on the observed density contrast value. The reliability of the inversion then depends on the reliability of the density contrast  measurement.

The advantage of the proposed technique with respect to \cite{Nakariakov01,GAA02, Aschwanden03b} is that it uses the information contained in both the period and the damping rate of oscillating coronal loops. In addition it makes no assumption on the particular value of the unknown parameters. The advantage with respect to the analytic and numerical inversions by \cite{Arregui07,GABW08} is that, first,  it enables us to  constrain  the transverse inhomogeneity length-scale and the density contrast. Second, the method incorporates  confidence levels and error bars consistently calculated  from the uncertainties on observed wave properties. Bayesian inference gives  a measure of degree of belief  about a proposition (set of parameters)  given some conditionals (observed data), once a probability model is set up. In that sense it is appropriate to perform inversions by making the fewer possible assumptions in the comparison to observations, as proposed by \cite{Arregui07} and \cite{GABW08}.  Our study is based on a theoretical model that assumes the classic straight magnetic flux tube representation for coronal loops and resonant absorption as damping mechanism. If any of these assumptions deviates from reality, then this will introduce errors into the inversion presented in this paper. For instance, \cite{demoortelpascoe09} have shown that the magnetic field strength inside a three-dimensional numerical coronal loop model may substantially differ from that inferred using classic seismology inversions in straight magnetic flux tubes.

The presented inversion technique can be applied to the determination of  physical parameters in other magnetic and plasma structures of the solar atmosphere that
display oscillatory dynamics.  One example are transversely oscillating prominence threads, provided a large part of the magnetic flux tube is filled with dense plasma \citep[see][]{Soler102dthread}. 
In this case, and because of the large density contrast typical of prominence plasmas, with internal densities two orders of magnitude larger than coronal densities, 
this parameter becomes irrelevant  to the inversion process \citep{AB10}. If we apply our algorithm to a case with 
$P=3$ min and $\tau_d=9$ min (with 10\% error), by considering large density contrast limits in the prior information for this parameter, we  find estimates of
$l/R=$0.22$^{+0.03}_{-0.03}$, $\tau_{\rm Ai}=$129$^{+12.8}_{-13.1}$, in excellent agreement with the analytic inversion approximation of  $l/R=0.21$, 
$\tau_{\rm Ai}=126.9$.  Although it may seem that in this case, as in the coronal loop inversion with Gaussian prior information, a two parameter-two observable problem 
seems to be fully constrained, we should be aware that we are dealing with incomplete information, because of uncertainties from measured quantities. The advantage of our method then lies in the consistent propagation of errors.

One of the most appealing applications of Bayes' rule is model comparison, where the evidence in Equation~(\ref{bayes}) plays the 
fundamental role. Provided different damping mechanisms could be properly parameterized with similar physical model parameters and observables, a 
Bayesian model comparison could be performed in order to shed light on the still highly debated mechanism that produces  the damping of 
transverse oscillations in coronal structures.

\acknowledgments
I.A. acknowledges the funding received under the project AYA2006-07637 by Spanish Ministerio de Ciencia e Innovaci\'on (MICINN) 
and FEDER Funds.  A.A.R. acknowledges the financial support of the Spanish Ministerio de Ciencia e Innovaci\'on (MICINN) through
project AYA2010--18029 (Solar Magnetism and Astrophysical Spectropolarimetry) and CONSOLIDER INGENIO CSD2009-00038 
(Molecular Astrophysics: The Herschel and Alma Era).  This work was motivated by discussions held at the II Reuni\'on Espa\~nola de 
F\'{\i}sica Solar y Heliosf\'erica in the framework of RIA (Red de Infraestructuras de Astronom\'{\i}a) and funded by the Direcci\'on 
General de Planificaci\'on y Coordinaci\'on of the Spanish MICINN, under the action CAC-2007-48. We are grateful to Jaume Terradas 
and Ram\'on Oliver for valuable comments. I.A. dedicates this paper to the memory of Ezequiel Hierro Palacio.


\clearpage


\clearpage



\clearpage


\clearpage


\clearpage


\begin{deluxetable}{cccccccccc} 
\tablecolumns{10} 
\tablewidth{0pc} 
\tablecaption{Inversion of synthetic data for different priors types and ranges of variation in density contrast. \label{table:synthetic}} 
\tablehead{ 
\colhead{Prior type}   & \multicolumn{2}{c}{Uniform} & \colhead{} &  \multicolumn{2}{c}{Jeffreys}  &\colhead{} & \multicolumn{3}{c}{Gaussian} \\ 
\cline{2-3}\cline{5-6}\cline{8-10} \\ 
\colhead{$\zeta$-range} & \colhead{$l/R$}    & \colhead{$\tau_{\rm Ai}$ (s)}  &\colhead{} &\colhead{$l/R$}    & \colhead{$\tau_{\rm Ai}$ (s)} & \colhead{} & \colhead{$\zeta$}   & \colhead{$l/R$}    & \colhead{$\tau_{\rm Ai}$ (s)}}
\startdata 
1.2--10  &  0.36$^{+0.70}_{-0.15}$&142.9$^{+19.0}_{-19.8}$   &    & 0.35$^{+0.23}_{-0.12}$& 142.9$^{+17.5}_{-16.7}$ & &4.99$^{+0.52}_{-0.53}$&0.26$^{+0.04}_{-0.04}$&152.4$^{+15.4}_{-16.2}$\\
1.2--15  &  0.26$^{+0.44}_{-0.09}$ &149.9$^{+18.6}_{-19.0}$   &    &0.31$^{+0.28}_{-0.12}$ & 144.8$^{+18.6}_{-18.6}$ & &4.95$^{+0.51}_{-0.51}$&0.26$^{+0.04}_{-0.04}$&152.1$^{+15.0}_{-15.4}$\\
1.2--20  &  0.22$^{+0.11}_{-0.05}$ &153.6$^{+16.6}_{-17.3}$   &    & 0.30$^{+0.22}_{-0.10}$& 147.2$^{+17.6}_{-18.9}$ & &4.99$^{+0.51}_{-0.50}$&0.25$^{+0.04}_{-0.04}$&151.1$^{+15.8}_{-16.2}$\\
1.2--30  &  0.21$^{+0.09}_{-0.05}$ &155.1$^{+18.5}_{-19.6}$   &    & 0.30$^{+0.33}_{-0.12}$& 146.4$^{+20.2}_{-21.0}$ & &4.97$^{+0.49}_{-0.52}$&0.25$^{+0.04}_{-0.03}$&152.0$^{+15.3}_{-15.4}$\\
1.2--40  &  0.19$^{+0.06}_{-0.04}$ &157.4$^{+17.6}_{-16.8}$   &    & 0.28$^{+0.22}_{-0.09}$& 146.1$^{+19.6}_{-17.7}$ & &4.99$^{+0.52}_{-0.52}$&0.25$^{+0.04}_{-0.04}$&152.0$^{+15.3}_{-15.9}$\\
1.2--50  &  0.19$^{+0.05}_{-0.03}$ &159.5$^{+17.7}_{-18.5}$   &    &0.25$^{+0.21}_{-0.08}$ & 150.4$^{+19.6}_{-23.5}$ & &4.98$^{+0.51}_{-0.51}$&0.26$^{+0.04}_{-0.04}$&151.5$^{+15.3}_{-15.3}$\\
1.2--60  &  0.19$^{+0.04}_{-0.03}$ &160.9$^{+17.8}_{-17.4}$   &    &0.24$^{+0.20}_{-0.08}$ & 151.3$^{+20.5}_{-20.5}$ & &5.05$^{+0.49}_{-0.51}$&0.25$^{+0.04}_{-0.04}$&151.7$^{+15.5}_{-14.8}$\\
1.2--100  &  0.19$^{+0.04}_{-0.03}$ &161.3$^{+17.3}_{-17.6}$   &    & 0.25$^{+0.23}_{-0.08}$& 155.1$^{+23.7}_{-22.2}$ & &4.98$^{+0.49}_{-0.48}$&0.26$^{+0.04}_{-0.04}$&151.9$^{+15.4}_{-14.9}$\\
\enddata 
\end{deluxetable}

\clearpage


\begin{deluxetable}{ccccrrrrrrrrr}
\tablecolumns{13} 
\tablewidth{0pc} 
\tablecaption{Analytic and Bayesian inversion results for the analyzed loop oscillation events.\label{restable} } 
\tablehead{ 
  \multicolumn{4}{c}{Oscillation properties} &   \colhead{}   & 
\multicolumn{8}{c}{Inversion results} \\ 
\cline{1-4} \cline{6-13} \\ 
\colhead{}&\colhead{}&\colhead{}&\colhead{}&\colhead{}&\colhead{Analytic}&\colhead{}&\multicolumn{2}{c}{Bayesian$_{\rm \; Jeffreys}$}&\colhead{}&\multicolumn{3}{c}{Bayesian$_{\rm \; Gaussian}$}\\
 \cline{6-6}\cline{8-9}\cline{11-13}\\ 
\colhead{\#} & \colhead{P (s)}   & \colhead{$\tau_{\rm d}$ (s)}    & \colhead{$P/\tau_{\rm d}$} & 
\colhead{}    & \colhead{$\tau_{\rm Ai}$ (s) }   & \colhead{}&\colhead{$\tau_{\rm Ai}$ (s)}    & \colhead{$l/R$} &\colhead{}&\colhead{$\tau_{\rm Ai}$ (s)}    & \colhead{$l/R$} &\colhead{$\zeta$}}
\startdata 
1  &  261 &  870 &  0.30  && 145--177 & &161.5$^{+22.2}_{-19.7}$&0.36$^{+0.27}_{-0.13}$&&169.4$^{+17.4}_{-16.9}$&0.30$^{+0.05}_{-0.04}$&4.99$^{+0.50}_{-0.50}$\\
2  &  265 &  300 &  0.88 &&  163--182  & &169.9$^{+20.9}_{-21.4}$&0.92$^{+0.47}_{-0.25}$&&167.1$^{+17.4}_{-16.9}$&1.01$^{+0.19}_{-0.16}$&3.76$^{+0.64}_{-0.61}$\\
3  &  316 &  500 &  0.63 && 189--217  & &199.4$^{+25.0}_{-24.5}$&0.76$^{+0.61}_{-0.28}$&&196.8$^{+17.4}_{-16.9}$&0.77$^{+0.43}_{-0.19}$&3.53$^{+1.88}_{-1.42}$\\
4  &  277 &  400 &  0.69 && 168--189  & &176.2$^{+22.7}_{-22.7}$&0.73$^{+0.53}_{-0.22}$&&167.2$^{+17.4}_{-16.9}$&1.05$^{+0.43}_{-0.28}$&2.56$^{+0.98}_{-0.69}$\\
5  &  272 &  849 &  0.32 &&  151--187  & &173.2$^{+21.1}_{-22.4}$&0.34$^{+0.26}_{-0.11}$&&159.7$^{+17.4}_{-16.9}$&0.58$^{+0.47}_{-0.17}$&2.18$^{+0.75}_{-0.62}$\\
6  &  522 & 1200 & 0.44 && 304--359 & &329.7$^{+43.8}_{-43.8}$&0.49$^{+0.39}_{-0.16}$&&319.9$^{+17.4}_{-16.9}$&0.59$^{+0.25}_{-0.13}$&2.97$^{+0.94}_{-0.91}$\\
7  &  435 &  600 &  0.73&& 267--299  & &281.3$^{+33.1}_{-35.4}$&0.74$^{+0.41}_{-0.20}$&&290.9$^{+17.4}_{-16.9}$&0.64$^{+0.11}_{-0.09}$&6.98$^{+1.05}_{-1.02}$\\
8  &  143 &  200 &  0.72&&  90--98  & &90.9$^{+12.0}_{-11.4}$&0.76$^{+0.53}_{-0.23}$&&93.8$^{+17.4}_{-16.9}$&0.69$^{+0.11}_{-0.10}$&5.55$^{+0.94}_{-0.96}$\\
9  &  423 &  800  & 0.53 &&247--291 & &  265.6$^{+35.2}_{-33.0}$&0.64$^{+0.68}_{-0.25}$&&290.5$^{+17.4}_{-16.9}$&0.41$^{+0.07}_{-0.06}$&13.4$^{+3.50}_{-3.80}$\\
10 & 185  &  200 &  0.93 &&117--126 & &119.2$^{+14.8}_{-14.8}$&0.94$^{+0.48}_{-0.26}$&&114.4$^{+17.4}_{-16.9}$&1.21$^{+0.24}_{-0.20}$&3.08$^{+0.43}_{-0.44}$\\
11 & 390 &  400 &  0.98 && 245--270   & &250.5$^{+29.6}_{-22.7}$&0.99$^{+0.54}_{-0.28}$&&221.5$^{+17.4}_{-16.9}$&1.69$^{+0.17}_{-0.25}$&2.10$^{+0.29}_{-0.23}$\\
\enddata 
\end{deluxetable} 

\end{document}